%% file: main.tex
\documentclass[reprint,
superscriptaddress,
longbibliography,
showpacs,
preprintnumbers,
nofootinbib,
nobibnotes,
amsmath,
amssymb, 
aps,
prr]{revtex4-1}
\input{packages}

\input{commands}
\input{Zallman.fd}
\begin{document}

\input{prelude}

\begin{abstract}

In the quest to unveil the laws of Nature, particle detectors produce an ever-increasing amount of data.
Large data rates require the use of specialized triggers that promptly reduce the data rate to a manageable level; however, in doing so, unexpected new phenomena may escape detection.
In this work, we present a methodology based on recent quantum compression techniques that can store exponentially more information than classically available methods.
We encode the full neutrino telescope event information using parity observables in an IBM quantum processor using 8 qubits to demonstrate this.
Then we show that we can recover the information stored on the quantum computer with a fidelity of 84\%. 
Finally, we illustrate the use of our protocol by performing a classification task that separates electron-neutrino events to muon-neutrinos events in a neutrino telescope.
We find that this classification method is limited by our ability to find adequately faithful representations of the classical data within the quantum systems, necessitating further investigations into finding these representations.
\end{abstract}

\maketitle

\input{introduction}

\input{po_encoding}
\input{demonstration}
\label{sec:conclusion}
\input{outlook}
\input{acknowledgements}

\bibliography{qencoding.bib}

\pagebreak
\clearpage
\onecolumngrid
\appendix

\ifx \standalonesupplemental\undefined
\setcounter{page}{1}
\setcounter{figure}{0}
\setcounter{table}{0}
\setcounter{equation}{0}
\fi

\renewcommand{\thepage}{Supplemental Methods and Tables -- S\arabic{page}}
\renewcommand{\figurename}{SUPPL. FIG.}
\renewcommand{\tablename}{SUPPL. TABLE}

\AtBeginEnvironment{appendices}{\crefalias{section}{appendix}}

\begin{appendices}

\input{app_data_generation}
\input{app-context-eigenstates}
\input{app-eff-parity}
\input{app-context-measurement}
\input{app-circuit-optimization}
\input{app-context-eigenstates-in-code.tex}
\input{app-collective-parity-cost-function.tex}
\input{app-state-optimization}
\end{appendices}

\end{document}

%% file: packages.tex
\usepackage{blindtext}
\usepackage[T1]{fontenc} 
\usepackage[utf8]{inputenc}
\usepackage[colorlinks=true,linkcolor=blue]{hyperref}%
\usepackage{graphicx}
\usepackage{url}
\usepackage{bm}
\usepackage{booktabs}
\usepackage{lineno}
\usepackage{ulem}
\usepackage{listings}
\usepackage[dvipsnames]{xcolor}
\usepackage[caption=false]{subfig}
\usepackage{tabularx}
\usepackage{lmodern}
\usepackage{gensymb}
\usepackage{fontawesome}
\usepackage{physics}
\usepackage{color}
\usepackage{float}
\usepackage{amsmath}
\usepackage{amsfonts}
\usepackage{amssymb}
\usepackage{relsize}
\usepackage{lettrine}
\usepackage{siunitx}
\usepackage[capitalise]{cleveref}
\usepackage{bbold}
\usepackage{appendix}
\usepackage{etoolbox}
\usepackage{pythonhighlight}
\usepackage{lipsum}

%% file: prelude.tex
\renewcommand{\LettrineFontHook}{\usefont{U}{Zallman}{xl}{n}}
\renewcommand{\LettrineTextFont}{\scshape}
\setcounter{DefaultLines}{3}%

\title{New Pathways in Neutrino Physics via Quantum-Encoded Data Analysis}

\author{Jeffrey~Lazar}
\email{jeffreylazar@fas.harvard.edu}
\affiliation{Department of Physics and Wisconsin IceCube Particle Astrophysics Center, University of Wisconsin–Madison, Madison, WI 53706, USA}
\affiliation{Department of Physics and Laboratory for Particle Physics and Cosmology, Harvard University, Cambridge, MA 02138, US}
\author{Santiago Giner Olavarrieta}
\email{santiagoginer@college.harvard.edu}
\affiliation{Department of Physics and Laboratory for Particle Physics and Cosmology, Harvard University, Cambridge, MA 02138, US}
\author{Giancarlo Gatti}
\email{gatti.gianc@gmail.com}
\affiliation{Department of Physical Chemistry, University of the Basque Country UPV/EHU, Apartado 644, 48080 Bilbao, Spain}
\affiliation{EHU Quantum Center, University of the Basque Country UPV/EHU, Apartado 644, 48080 Bilbao, Spain}
\affiliation{Quantum Mads, Uribitarte Kalea 6, 48001 Bilbao, Spain}
\author{Carlos A. Arg\"{u}elles}
\email{carguelles@fas.harvard.edu}
\affiliation{Department of Physics and Laboratory for Particle Physics and Cosmology, Harvard University, Cambridge, MA 02138, US}
\author{Mikel Sanz}
\email{mikel.sanz@ehu.es}
\affiliation{Department of Physical Chemistry, University of the Basque Country UPV/EHU, Apartado 644, 48080 Bilbao, Spain}
\affiliation{EHU Quantum Center, University of the Basque Country UPV/EHU, Apartado 644, 48080 Bilbao, Spain}
\affiliation{IKERBASQUE, Basque Foundation for Science, Plaza Euskadi 5, 48009, Bilbao, Spain}
\affiliation{Basque Center for Applied Mathematics (BCAM),
Alameda de Mazarredo, 14, 48009 Bilbao, Spain}




%% file: introduction.tex
\section{Introduction}
\label{sec:intro}

\begin{figure*}[ht!]
\centering
    \includegraphics[width=0.95\textwidth]{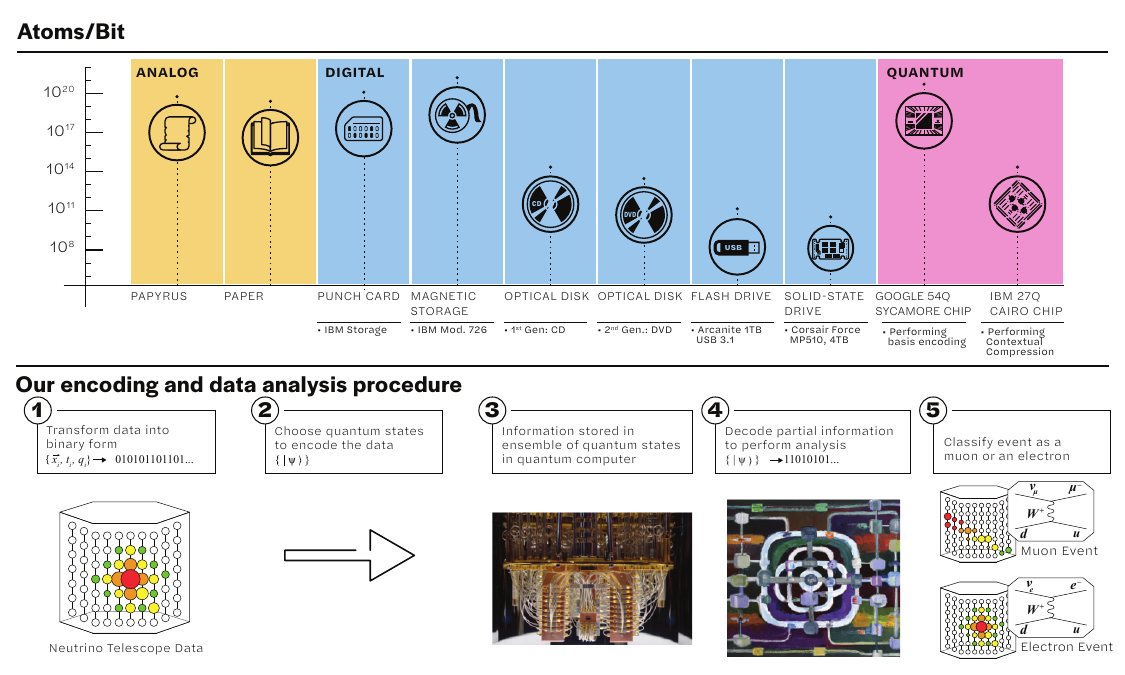}
\caption{\textbf{\textit{Data compression through history.}}
The vertical axis shows the number of atoms needed to encode a single bit of information, while the horizontal axis shows the technology or medium used to store it.
We calculated the number of atoms per bit in analog formats (shown in yellow) by counting the average number of characters per document in Greek and English, respectively.
For digital medium (shown in blue), we estimated the number of atoms from typical ISO fabrication standards.
For quantum devices (shown in pink), the number of bits accessible from $54$-qubit quantum states in Google's Sycamore chip is first calculated for a straightforward \textit{basis encoding} (left) and afterwards calculated in a \textit{contextual quantum random access code} (right), extrapolating the method used in this work into that scale.
The bottom section shows a high-level overview of this work, starting with the simulation of a high-energy neutrino interaction to the classification of the decoded data.
Intermediate steps are represented by the IBM Q quantum computer and an artistic interpretation of quantum computation.
}
\label{fig:overview}
\end{figure*}

Since the Geiger-Muller counter was used to trigger on $\alpha$ particles in 1928, automated triggering systems have become increasingly important to experimental high-energy physics~\cite{Galison1997-GALIAL-2}.
The increased need for higher intensity and higher energy beams resulted in the development of faster electronics and complementary data storage and analysis techniques.
In particular, the development of magnetic storage on tapes, disks, and solid state drives (see~\cref{fig:overview} (A)) enabled the \textit{Big Data} paradigm of particle physics, culminating in the 2012 discovery of the Higgs Boson at the Large Hadron Collider (LHC)~\cite{ATLAS:2012yve,CMS:2012qbp}.
The lockstep development of the Standard Model (SM) and novel data storage techniques echoes the historical precedent, where efficient data storage and retrieval has been central to advancing our understanding of Nature.

While the successes of our current paradigm are undeniable~\cite{Workman:2022ynf}, significant questions remain unanswered on both the largest and the smallest scales. For example, on galactic and cosmological scales, the SM cannot provide a viable dark matter candidate.
Furthermore, at the sub-nuclear scale, the origin and smallness of neutrino masses continue to elude us. In an attempt to answer these questions, high-energy physics (HEP) experiments use triggers inspired by familiar physics models to select and reduce the data to a manageable level; however, this approach leaves us vulnerable to the so-called streetlight effect, an observational bias which leads us to only search in the best known areas.
As we continue to pursue new physics, we may need to relax these filters to allow in previously unconsidered types of events.

This need to expand our searches to new regions is happening against the backdrop of a crisis of data management and acquisition in HEP.
For example, every day, the IceCube Neutrino Observatory produces one Terabyte (TB) of data~\cite{IceCube:2016zyt}, while the LHC produces nearly 300~TB in the same period~\cite{Arsuaga-Rios:2021lbo}.
Additionally, both experiments have planned upgrades~\cite{IceCube-Gen2:2020qha,Apollinari:2015wtw} that will increase their data production by an order of magnitude, IceCube by expanding its volume and the LHC by increasing its luminosity.
Beyond these two experiments, many other large volume next-generation experiments such as DUNE~\cite{DUNE:2020lwj} and HyperKamiokande~\cite{Hyper-Kamiokande:2018ofw} are expected to face similar challenges.
While it may be possible to meet these data needs using current technologies, selecting and storing large fractions of the data will become increasingly untenable,
rendering us more vulnerable to the streetlight effect.

Quantum technologies make use of degrees of freedom provided by quantum physics to enhance the performance in some classical tasks.
For instance, in quantum metrology these additional resources improve the accuracy of sensors, pushing it towards the Heisenberg limit~\cite{Degen_2017}.
Additionally, quantum communication designs secure communication protocols against a possible eavesdropper~\cite{gisin2007quantum}.
Finally, resources such as entanglement and superposition are commonly exploited by quantum computation to speed up information processing tasks~\cite{monroe2002quantum}.


%% file: po_encoding.tex
\section{Encoding Protocol and Implementation}
\label{sec:protocol}

Our goal is to encode a classical \textit{target bitstring}---a sequence of 0s and 1s---in a system of qubits efficiently.
A bitstring is a natural choice for quantum storage since qubits are two-level systems, meaning that any measurement of a qubit's spin will yield only $\lambda_{\pm} = \pm 1$.
These outcomes may be straightforwardly mapped to a bit value via ${\rm{bit}}(\lambda)=(1-\lambda)/2$; however, encoding one classical bit per qubit is inefficient.
We may extend this thinking and define the \textit{parity} of a many-qubit system as the product of spin measurements on each qubit.
Since the measurement of each qubit will yield $\pm 1$, the product of all measurements will also yield $\pm 1$.
Furthermore, since we may measure the spin along either the $x$-, $y$-, or $z$-direction, we can construct $3^n$ \textit{parity operators} (POs) for an $n$-qubit system.

\begin{figure*}[t!]
\includegraphics[width=0.9\textwidth]{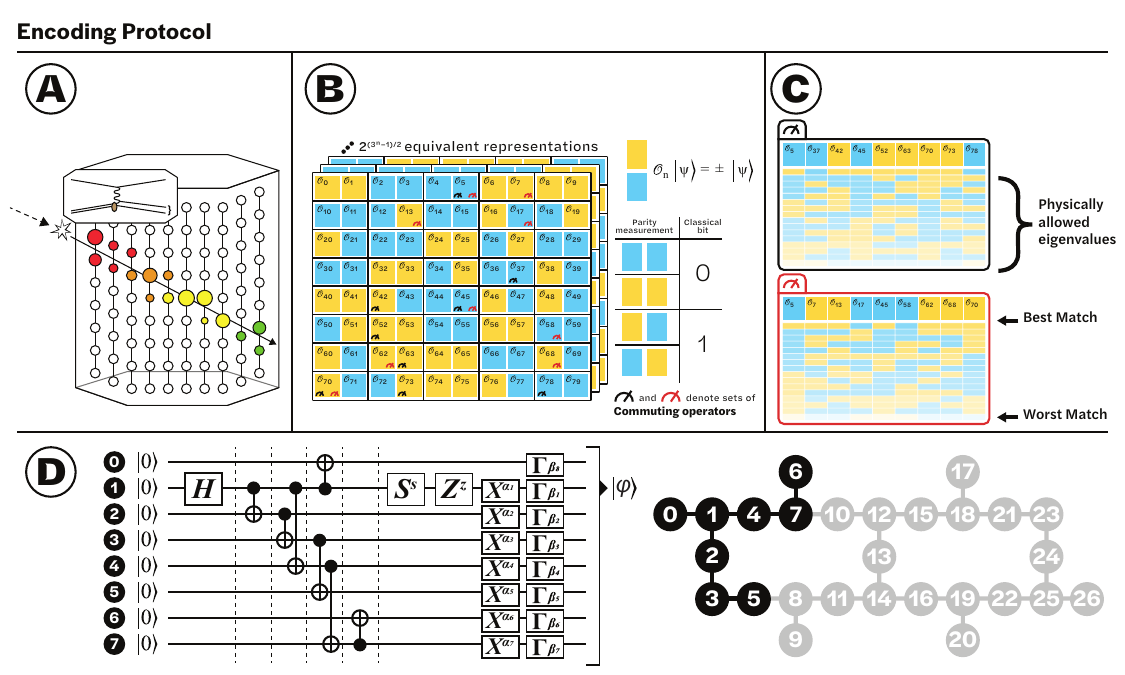}
\caption{\textbf{\textit{Simulation schematic and encoding protocol.}} 
A.: Illustration of a neutrino interaction in a neutrino telescope.
The colored circles highlight the activated sensors, while the color scale is correlated with time.
B.: Visual representation of $80$ parity measurements (out of the $81$ parity observables $\{X,Y,Z\}^{\otimes 4}$) performed in a selection of $4$-qubit quantum states.
Each observable has a majority parity of either $+1$ or $-1$ in the state selection, and the observables are grouped in pairs, such that a pair with two equal majority parities represents classical bit $0$, and a pair with two different majority parities represents classical bit $1$.
This sequence of $80$ observables represents the bitstring $01000 01011 10010...010$ of length $40$.
In general, if $n$-qubit systems are considered, there are $(3^n-1)/2$ observable pairs, and each bitstring can be represented by $2^{(3^n-1)/2}$ different configurations of majority parities.
Additionally, the measurement symbols demark two sets of commuting observables.
C.: Top row corresponds to the desired parity configuration of the set of commuting observables. 
Narrow, lightly-colored rows below correspond to physically accesible configuration order from best-match (top) to worst (bottom).
D.: Circuit that produces parity eigenstates within different CSCOs.
The sub-set of the qubits in IBM-Cairo used in this work are highlighted in black.
}
\label{fig:encoding}
\end{figure*}

With these POs, we can recover a target bitstring of length $3^n$ by measuring them on an ensemble of $n$-qubit systems.
Formally, we choose a mapping, $\mathcal{O}$, between the non-negative integers less than $3^n$, $\left\{x\in \mathbb{Z} | 0\leq x <3^n \right\}$, and the set of all POs, $\left\{X,\, Y,\, Z \right\}^{\otimes\,n}$.
This mapping does not play a crucial role in our algorithm and may be chosen arbitrarily.
This allows us to define a mapping between a set of $n$-qubit systems and the target bitstring, $b$ of length $3^n$ whose values are indexed by $i$, via:  

\begin{equation}
\label{eq:round}
    b_{i} = {\rm{round}}\left[\text{bit}\bigg(\frac{1}{\left|\left\{\left|\phi\right>\right\}\right|}\sum_{j}\left<\phi_{j}\right|\mathcal{O}(i)\left|\phi_{j}\right>\bigg)\right]\mathrm{,}
\end{equation}
where $\left\{\left|\phi\right>\right\}$ is the set of $n$-qubit states, and $\left|\phi_{j}\right>$ is a specific state within that set.
Simply put, to compute the $i$\textsuperscript{th} bit, we take a classical average of the expectation values for $\mathcal{O}(i)$ over all states---mapped to a value between $0$ and $1$ via the $\text{bit}(\cdot)$ function---and then round this value.
Thus, to store a given bitstring in a set of $n$-qubit systems, we need only find a set of quantum states which faithfully reproduce this bitstring under the prescription given in~\cref{eq:round}.

However, directly encoding classical bits to rounded PO measurements in a one-to-one map presents a challenge.
Since quantum mechanics restricts the relationships between eigenvalues for certain spin measurements, it may not be possible to succinctly represent an arbitrary set of bits with a set of quantum states in this manner.
The encoding procedure described in Ref.~\cite{Gatti:2021aou}, circumvents this challenge by encoding the bits into pairs of PO measurements, rather than mapping each bit to a particular PO measurement.

Under this procedure, the value of the target bitstring is obtained by applying the \texttt{XOR} operation to pairs of quantum measurements, giving 0 if the measurements are equal and 1 otherwise; see the inlay of~\cref{fig:encoding} (B).
While this choice halves the number of classical bits that can be encoded in $n$-qubits, this achieves immense freedom since now a target bitstring of length $N$ can be represented by $2^{N}$ equivalent \textit{representations}, some of which may be more convenient than others.
This equivalency is represented by the three stacked grids in~\cref{fig:encoding}~(B).

A good representations of this data will be readily expressed by physically allowed PO eigenstates while using minimal quantum resources.
To meet the latter requirement, we use eigenstates of complete sets of commuting observables (CSCOs), which have a size of $2^{n-1}+1$~\cite{Gatti:2021aou}.
This ensures we can extract the maximal amount of information from a single quantum state.
We construct such states starting from the base CSCOs:
\begin{equation}
\label{eq:context_0}
    \mathcal{C}^{\text{even}}_0 = \{Z^{\otimes n}\}\cup\{X,Y\}^{\otimes n}_{\text{even number of $X$s}},
\end{equation}
and
\begin{equation}
\label{eq:context_1}
    \mathcal{C}^{\text{odd}}_0 = \{Z^{\otimes n}\}\cup\{X,Y\}^{\otimes n}_{\text{odd number of $X$s}}.
\end{equation}
%

We note three important facts about these sets of operators.
First, all operators in each set mutually commute, which is is easy to check from known anti-commutation relations that.
Second, the size of each set is $1 + 2^{n} / 2 = 1 + 2^{n-1}$, where the division by two arises from the requirement that there be an even or odd number of $X$ operators.
Together, these two factors show that these definitions do indeed define a CSCO.
Finally, the same anti-commutation relation reveals that the only operator that mutually commutes between the two CSCOs is $Z^{\otimes n}$.
Thus, we say that $Z^{\otimes n}$ generates these two CSCOs.

We can then create new CSCOs from the base CSCOs by applying rotations of the form:
\begin{equation}
\label{eq:context_rotation}
    R(\vec{\beta}) = \Gamma_{\beta_{1}}^{1}...\Gamma_{\beta_{n}}^{n},
\end{equation}
where the subscript $\beta_{x}\in \{0,1,2\}$ and the $\Gamma$ operators are given by:
\begin{align}
\label{eq:beta_defs}
    \Gamma_{0} &= \mathbb{1} \\
    \Gamma_{1} &= H\,Z\,S \\
    \Gamma_{2} &= S\,H. 
\end{align}
Here, $H$ is the Hadamard gate and $S$ is the phase gate.
It is straightforward to show that, up to a phase, $\Gamma_{1}$ cyclically permutes the $X$, $Y$, and $Z$ eigenstates, while $\Gamma_{2}$ cyclically permutes the $X$, $Z$, and $Y$ eigenstates.
See App~\ref{appendix:PO_eigenstates} for further details.

Since there are $3^{n}$ possible $R$, there are $2 \times 3^{n}$ CSCOs.
Furthermore, each CSCO contains $2^{n-1}+1$ parity observables and can take on $2^{n}$ eigenvalue configurations~\cite{Gatti:2021aou}.
It is not computationally tractable to compute all $2 \times 3^{n} \times (2^{n-1} +1) \times 2^n\sim12^n$ eigenvalues for any reasonably large value of $n$; however, we can exploit the fact that the $\Gamma$ operators preserve the eigenvalue.
Thus, we need only calculate a much more manageable $2 \times (2^{n-1} +1) \times 2^{n}\sim 4^n$ eigenvalues for the base CSCOs, and then---provided we are careful about ordering the POs correctly---we may use these across all CSCOs.
We further accelerate this process by via an analytic expression we computed for the parity of a state in terms $\vec{\beta}$.
See App.~\ref{appendix:fingerprint} for further details about this expression.

Armed with these computational improvements, we turn our attention to evaluating the fitness of a representation.
To do this, we begin by picking a CSCO and iterating over the $2\times3^{n}$ possible CSCOs to find the physically allowed eigenstate whose eigenvalues minimize the Hamming distance between eigenvalues and the sub-portion of the representation corresponding to the CSCO POs~\cite{Gatti:2024thesis}; see~\cref{fig:encoding} (C) for examples of this comparison.
We then sum the Hamming distances taken over all CSCOs to compute a heuristic for the ease with which the representations can be expressed by the physically allowed CSCO eigenstates.

In order to perform and optimization, we parameterize a representation as yet another bitstring, with length $(3^n - 1) / 2$.
In this parameterization, the bit determines the first bit of each bit pair in the representation and the corresponding bit in the target bitstring determines the second bit in each pair via the \texttt{XOR} convention.
We may then use well-established methods in genetic algorithms~\cite{banzhof1998} to optimize the bitstring-parameterized representation, where the total Hamming distance is the objective function to be minimized.
As we go, we also record the eigenvalues that were closest to the appropriate subsection of the representation for each CSCO.

The final step for encoding the target bitstring is selecting a subset of the CSCOs from the previous step that can faithfully reproduce the desired representations.
Once again, this can be parameterized in terms of a bitstring of length $2\times 3^{n}$, where each bit corresponds to whether or not a particular CSCO will be used.
Thus, we may apply the machinery of genetic algorithms to perform this optimization.
To do this, we use a heuristic function that balances the faithfulness of the representations with the amount of quantum resources used.
This heavily penalizes sets of CSCOs that do not faithfully reproduce the desired representation, while encouraging the use of fewer quantum resources.

We may then encode these values onto a quantum backend.
To do this, we use a circuit of the form show in \cref{fig:encoding} (D).
We first generate a Greenberger-Horne-Zeilinger (GHZ) state by applying a Hadamard gate followed by a series of controlled not gates.
This creates a state of the form, e.g.:
$$
\left|\phi\right> = \frac{1}{\sqrt{2}}\big[\left|0000\right> + \left|1111\right>\big]
$$
We then modify the relative phase by applying two bit-parameterized gates, $S^{*}(b_{s})$ and $Z^{*}(b_{z})$, where each gate is the identity if the bit equals 0 and is $S$ or $Z$ gate if the bit equals 1.
This results in a state of the form:
$$
\left|\phi\right> = \frac{1}{\sqrt{2}}\big[\left|0000\right> + i^{2b_{z} + b_{s}} \left|1111\right>\big]
$$
We then apply $n-1$ bit-parameterized $X^{*}(\alpha_{i})$ gates, which is similarly defined as being equal to the identity if the bit equals 0 and $X$ if the bit equals 1.
This will exchange the value of 0 and one 1 to produce a state of the form, e.g.:
$$
\left|\phi\right> = \frac{1}{\sqrt{2}}\bigg[\left|1^{\alpha_{0}}1^{\alpha_{1}}1^{\alpha_{2}}0\right> + i^{2b_{z} + b_{s}} \left|1^{1-\alpha_{0}}1^{1-\alpha_{1}}1^{1-\alpha_{2}}0\right>\bigg]
$$
It is not important which qubit the $S^{*}$ and $Z^{*}$ gates are applied to as long as it is not the sole qubit that does not have a $X^{*}$ gate applied.
Finally, a set of $n$ $\beta$ gates, previously defined in \cref{eq:beta_defs}, is applied to each qubit.
This changes the eigenstate from the $z$-basis to the $x$—or $y$-basis, depending on the value of the gate parameter.

\begin{figure*}[t]
\includegraphics[width=\textwidth]{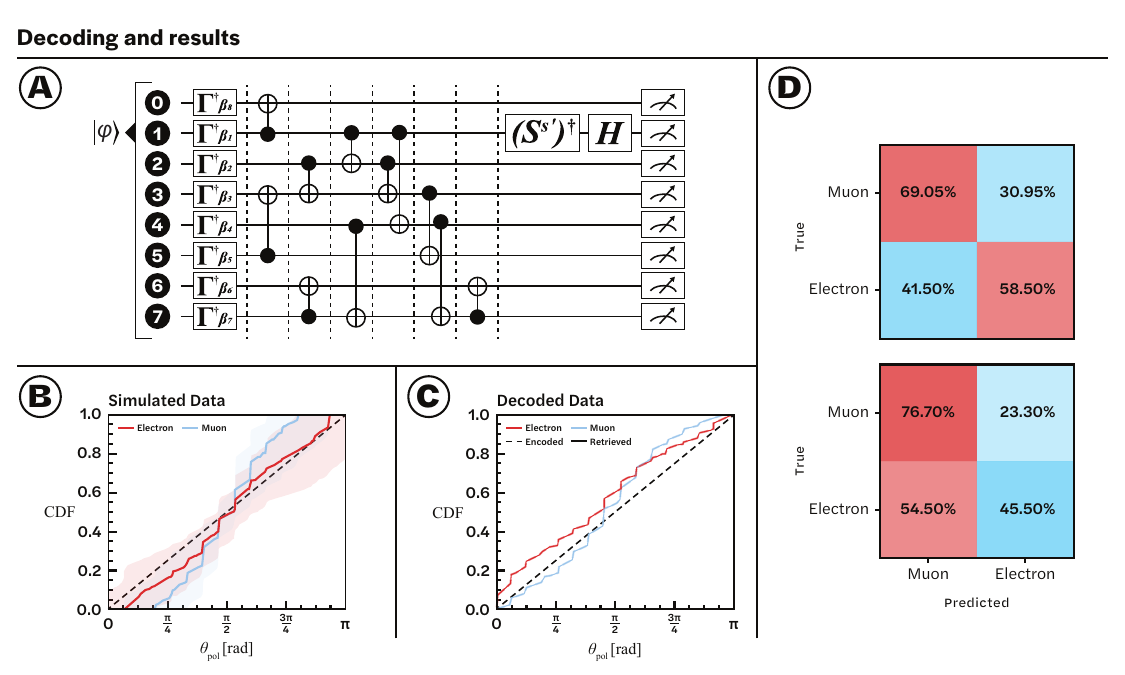}
\caption{\textbf{\textit{Decoding protocol and inference results.}}
A.: Shows the circuit used to measure the parity observables.
B.: Cumulative distribution of light sensor hits as a function angle. 
The red line corresponds to electron events, which produce approximately spherical light depositions in neutrino telescopes, while the blue light corresponds to muons, which make long tracks in the array. 
The shaded regions visually represent the event-by-event variability of these distributions for each category.
C.: The same distribution as in B, but for the data encoded and retrieved from the quantum computer in dashed and solid respectively.
Note that the encoded and retrieved data are almost identical, corresponding to the high fidelity of reading out information from the quantum backend.
However, the shapes of these distributions differ significantly from the simulated data, pointing to challenges in expressing the simulated information in quantum states.
D.: The confusion matrices for our particle type prediction.
This was constructed by placing straight cuts on the results of a Kolmogorov–Smirnov test applied to the distributions of the polar angles of OMs that saw light---\textit{i.e.} the information shown in sections B and C---for each event.
The top matrix is for the training set of simulated data, whereas the bottom matrix is for the test set data read out from the quantum backend.
The moderate success in differentiating events on the training set is lost by the errors when finding quantum states to encode the data.
}
\label{fig:decoding}
\end{figure*}

It is worth explicitly stating the relation between these circuit parameters and the theoretical procedure.
The value of the $b_{s}$ determines whether the CSCO is even or odd, i.e. whether you are generating an eigenstate according to \cref{eq:context_0} or \cref{eq:context_1}.
The values of the $\beta_{s}$ parameters define the rotation away from base CSCOs, i.e. \cref{eq:context_0} or \cref{eq:context_1}.
In fact, these are the same beta values as appear in \cref{eq:context_rotation}.
Finally, $b_{z}$ and $\alpha_{i}$ determine which set of eigenvalues will be used for the chosen CSCO.
All told, we have $2 \times 3^{n} \times 2^{n-1}$ possible choices of circuit.
When combined with the $2^{n+1}-1$ POs in each CSCO, we recover the expected $2 \times 3^{n} \times (2^{n-1} +1) \times 2^n$ different PO eigenvalues.

%% file: demonstration.tex
\section{Demonstration on Neutrino Telescope Data}
\label{sec:demo}

To demonstrate the encoding protocol in the context of high-energy physics, we simulate high-energy neutrino interactions in an ice-based neutrino telescope with a geometry inspired by those of current and future neutrino observatories~\cite{IceCube:2016zyt,IceCube-Gen2:2020qha}.
See \cref{fig:encoding} for a schematic representation of such geometry.
These observatories are comprised of thousands of light-detecting optical modules (OMs), which, at the most basic level, are photomultiplier tubes and digitization hardware inside a pressure housing.
These OMs detect the light created by the charged byproducts of neutrino interactions.
Furthermore, these OMs are arranged on strings of $\sim60$ OMs with the inter-OM and interstring spacing depending on the energy being targeted and the medium in which the observatory is embedded.
When targeting high-energy, cosmic neutrinos in ice, the interstring spacing is typically $\gtrapprox 150\,\mathrm{m}$

The pattern of deposited light will vary depending on the initial neutrino flavor and interaction type.
The two most common patterns, or \textit{morphologies}, are \textit{tracks} and \textit{cascades}.
The track morphology is characterized by long, linear depositions of light created by muons, which can travel several to tens of kilometers at the energies relevant to neutrino observatories~\cite{Koehne:2013gpa}.
These muons can be created by muon-neutrino charged-current interactions.
The cascade morphology, on the other hand, is characterized by a roughly spherical light emission profile.
These result from electron-neutrino charged-current interactions and all-flavor neutral current interactions when the charged byproducts deposit their energy within several to ten meters.
Since this is much smaller than the interstring spacing, the light emission profile appears roughly point-like to the detector.

In order to test whether the quantum encoded data could be used to distinguish physically relevant information, we simulated muon-neutrino and electron-neutrino charged current events using the \texttt{Prometheus} software.
In total, we simulated events with energies between $100\,\mathrm{GeV}$ and $5\times 10^{5}\,\mathrm{GeV}$ sampled according to a power law with a spectral index of 1.
We then selected events that resulted in at least 20 photons being observed in at most 20 OMs.
The former criterion ensures that events are of reasonable quality, while the second criterion is necessary to ensure that the events can be encoded in eight qubits via our protocol.

For the remaining events, we generate a bitstring that summarizes the information from the individual OMs.
In particular, we encoded the total number of photons that each OM saw, the average arrival time of each photon, and the position of each OM in a coordinate system center at the charge-weighted center of the event.
For reasons that will be discussed later, we encoded the position of each OM in spherical coordinates.
We then create our target bitstring by converting each variable to binary and concatenating the values for each OM that detected light.
We encoded $(3^8 -1) / 2=3,280$ bits of information containing the data for each of the approximately thousand events generated.
If an event had less than twenty triggered optical modules, we zero-pad the bit string to achieve the required length.

We then performed the encoding optimization described in \cref{sec:protocol} to embed each event into a quantum computer.
On median, we encoded the 3,280 bits of classical data into $680^{+18}_{-25}$ eight-qubit states, corresponding to $5440^{+144}_{-200}$ two-level systems.
Since the number of two-level states exceeds the number of classical bits, we did not achieve compression with this protocol.
This is not surprising as compression is not expected until at least 14 qubits~\cite{Gatti:2021aou}.
The optimized encoded states reproduced the target bitstring with $84.32\%^{+0.69\%}_{-1.08\%}$ accuracy.

We then embedded these states on the IBM-Q Cairo quantum processor and read out the parity with the circuit shown in \cref{fig:decoding} (A).
Our method recovered the theoretical state values with $100.00\%^{+0.00\%}_{-1.04\%}$ fidelity due to the low-depth circuit used to implement and measure GHZ states.

Finally, we tested whether we could use the encoded data to measure physically relevant properties.
In particular, we set out to differentiate tracks and cascades by looking at the distribution of polar angles among triggered OMs.
We then used half of the events to design cuts to distinguish these events based on the cumulative distribution functions (CDFs) of the unencoded polar angles, see \cref{fig:decoding} (B).
These cuts had moderate predictive success of 63.8\%.
We then applied the same cuts to the CDFs---see \cref{fig:decoding} (C)---from the decoded data and found that the predictive success had greatly decreased.
This seems to be due to systematic errors in the encoded events since the decoded and encoded distributions match quite well.


%% file: outlook.tex
\section{Conclusions and Outlook}

In this article, we have demonstrated the encoding of high-energy physics data into a quantum stream. 
The protocol introduced in this work can serialize exponential amounts of data and achieves greater compression than classical resources when more than fourteen qubits are used.
The availability of quantum resources in the upcoming years together with the techniques introduced in this work opens the door to trigger-less data analyses on high-energy physics.

Upcoming quantum computers such as the IBM-Q, Google Sycamore processor~\cite{Arute_2019}, or Rydberg quantum simulators~\cite{Ebadi_2021} can be used in the upcoming experimental analyses.
Our letter provides a new path forward to use quantum computers in high-energy physics data analyses~\cite{Delgado:2022tpc}.

%% file: acknowledgements.tex
\section*{Acknowledgments}
We would like to acknowledge useful discussions with Tim Green, Joseph Farchione, and Joseph Lazar.
Also we would like to thank Jerry Ling for their willingness to share expertise on the \texttt{Julia} language.
Kathy and Jerome Lazar for their hospitality that let to the inception of this project.
JL is supported by NSF under grants PLR-1600823 and PHY-1607644,by the University of Wisconsin Research Council with funds granted by the Wisconsin Alumni Research Foundation, and by the Belgian American Educational Foundation.
SG acknowledges support from the Harvard College Research Program in the fall of 2021.
CAA are supported by the Faculty of Arts and Sciences of Harvard University, the National Science Foundation, and the David \& Lucile Packard Foundation.
CAA and JL were supported by the Alfred P. Sloan Foundation for part of this work.
MS and GG acknowledge support from EU FET Open project EPIQUS (899368) and HORIZON-CL4- 2022- QUANTUM01-SGA project 101113946 OpenSuperQ- Plus100 of the EU Flagship on Quantum Technologies, the Spanish Ram\'on y Cajal Grant RYC-2020-030503-I, project Grant No. PID2021-125823NA-I00 funded by MCIN/AEI/10.13039/501100011033 and by ``ERDF A way of making Europe'' and ``ERDF Invest in your Future'', and from the IKUR Strategy under the collaboration agreement between Ikerbasque Foundation and BCAM on behalf of the Department of Education of the Basque Government. This project has also received support from the Spanish Ministry of Economic Affairs and Digital Transformation through the QUANTUM ENIA project call - Quantum Spain, and by the EU through the Recovery, Transformation and Resilience Plan - NextGenerationEU. 
We acknowledge the use of IBM Quantum services for this work.
The views expressed are those of the authors, and do not reflect the official policy or position of IBM or the IBM Quantum team.

%% file: app_data_generation.tex
\section{Data Generation}
\label{appendix:generation}

In this work, we have simulated a neutrino telescope with a geometry that approximates the one of the IceCube Neutrino Observatory to exemplify our encoding scheme.
The simulation chain uses the \texttt{Prometheus}~\cite{Lazar:2023rol} framework.
\texttt{PROMETHEUS} generates neutrino interactions around the detector volume by means of \texttt{LeptonInjector}~\cite{IceCube:2020tcq} by sampling from deep-inelastic cross section tables~\cite{Cooper-Sarkar:2011jtt}.
For charged-current interactions this produces an outgoing charged particle and a hadronic cascade, while for neutral current interaction an out-going neutrino is produced with an associated hadronic cascade.
In this work, we will only encode charge-current neutrino interactions, though our protocol can easily be extended for neutral currents interactions.
\texttt{Prometheus} uses \texttt{PROPOSAL}~\cite{Koehne:2013gpa} to simulate charge lepton energy losses in the detector media and an internal \texttt{PROMETHEUS} module to compute the corresponding hadronic cascade light yield.
The photons are then transported to the detector photomultipliers using the Photon Propagation Code.
Finally, the the time series of photons that reach each of the sensors is recorded.

Since the number of photons can be large for sensors near the detector, we store the total number of photons that reach the detector and the mean arrival time.
Thus our simulated data set consists of tuples of $\{x,y,z,q_{tot},\bar t\}$ where $q_{tot}$ is the total number of photons that arrived in the detector and $\bar t$ the mean time.

%% file: app-context-eigenstates.tex
\section{Context eigenstates}
\label{appendix:PO_eigenstates}

The procedure in this work encodes classical bits in the statistics of $n$-qubit parity observables (POs) $\{X,Y,Z\}^{\otimes n}$, also known as $n$-body Pauli observables.
To do this, eigenstates of complete sets of commuting observables (CSCOs) are employed.
Assuming even $n$, all CSCOs considered are rotations of the following CSCOs:
\begin{equation}
    \mathcal{C}^{\text{even}}_0 = \{Z^{\otimes n}\}\cup\{X,Y\}^{\otimes n}_{\text{even number of $X$s}},
\end{equation}
and
\begin{equation}
    \mathcal{C}^{\text{odd}}_0 = \{Z^{\otimes n}\}\cup\{X,Y\}^{\otimes n}_{\text{odd number of $X$s}}.
\end{equation}
Note that $\mathcal{C}^{\text{even}}_{0}$ and $\mathcal{C}^{\text{odd}}_{0}$ are comprised of $2^{n-1}+1$ elements each.
We choose an arbitrary order for these elements, setting $Z^{\otimes n}$ as the first element, followed by $\{X,Y\}^{\otimes n}$, with even or odd number of Xs, in alphabetical order. They are given by: $\mathcal{C}^{\text{even}}_{0}=\{Z^{\otimes n}\, ,\,X^{\otimes n}\, ,\,X^{\otimes (n-2)}\! \otimes \! Y \! \otimes \! Y\, , \,\text{...}\, , \,Y^{\otimes n}\}$ and $\mathcal{C}^{\text{odd}}_{0}=\{Z^{\otimes n}\, , \, X^{\otimes (n-1)}\! \otimes \! Y\, , \,X^{\otimes (n-2)} \! \otimes \! Y \! \otimes \! X\, , \,\text{...}\, , \,Y^{\otimes (n-1)}\! \otimes \! X\}$.
This way, we can refer to a specific observable $\mathcal{C}^{\text{even}}_{0}(i)$ or $\mathcal{C}^{\text{odd}}_{0}(i)$ of the CSCOs using an integer index $i$ with $0\le i \le 2^{n-1}$.

To define the class of rotations to apply on our CSCOs, we first define the Pauli $X$ eigenstates as $\ket{+}=\tfrac{1}{\sqrt{2}}(\ket{0}+\ket{1})$ and $\ket{-}=\tfrac{1}{\sqrt{2}}(\ket{0}-\ket{1})$, the Pauli $Y$ eigenstates as $\ket{L}=\tfrac{1}{\sqrt{2}}(\ket{0}+i \ket{1})$ and $\ket{R}=\tfrac{1}{\sqrt{2}}(\ket{0}-i \ket{1})$, and the Pauli $Z$ eigenstates as $\ket{0}$ and $\ket{1}$. 
Then, we define the Hadamard gate, $H = \tfrac{1}{\sqrt{2}}\bigl( \begin{smallmatrix}1 & 1\\ 1 & -1\end{smallmatrix}\bigr)$, and the Phase gate, $S = \bigl( \begin{smallmatrix}1 & 0\\ 0 & i\end{smallmatrix}\bigr)$.
Once this is done, we consider three basic rotations that can be used to map Pauli eigenstates to Pauli eigenstates, namely:
\begin{equation}
\label{eq:pauli_to_pauli}
    \Gamma_0 = \mathbb{1}\,\, , \,\,\Gamma_1=H\,Z\,S\,\, , \,\,\Gamma_2=S\,H,
\end{equation}
such that $\Gamma_0$ does nothing when applied to a quantum state, $\Gamma_1$ rotates Pauli eigenstates clockwise save for a global phase.
Explicitly:
\begin{eqnarray*}
\label{eq:paulieigen_to_paulieigen}
\text{$X$ eigenstates to $Y$ eigenstates}\\
\Gamma_1 \ket{+} &=& e^{- i \pi/4}\ket{L}, \\ 
\Gamma_1 \ket{-} &=& e^{+ i \pi/4}\ket{R}, \\
\\
\text{$Y$ eigenstates to $Z$ eigenstates}\\
\Gamma_1 \ket{L} &=& \ket{0}, \\ 
\Gamma_1 \ket{R} &=& \ket{1}, \\
\\
\text{$Z$ eigenstates to $X$ eigenstates}\\
\Gamma_1 \ket{0} &=& \ket{+}, \\ 
\Gamma_1 \ket{1} &=& e^{- i \pi/2}\ket{-},
\end{eqnarray*}
and $\Gamma_2=\Gamma^\dagger_1$ rotates Pauli eigenstates counterclockwise save for a global phase.
Explictly:
\begin{eqnarray*}
\label{eq:paulieigen_to_paulieigen2}
\text{$X$ eigenstates to $Z$ eigenstates}\\
\Gamma_2 \ket{+} &=& \ket{0}, \\ 
\Gamma_2 \ket{-} &=& e^{+ i \pi/2}\ket{1}, \\
\\
\text{$Y$ eigenstates to $X$ eigenstates}\\
\Gamma_2 \ket{L} &=& e^{+ i \pi/4}\ket{+}, \\ 
\Gamma_2 \ket{R} &=& e^{- i \pi/4}\ket{-}, \\
\\
\text{$Z$ eigenstates to $Y$ eigenstates}\\
\Gamma_2 \ket{0} &=& \ket{L}, \\ 
\Gamma_2 \ket{1} &=& \ket{R}.
\end{eqnarray*}
Furthermore, $\Gamma_1$ can be used to define a basis transformation to rotate Pauli observables clockwise, i.e.,
\begin{eqnarray*}
\label{eq:pauliop_to_pauliop}
\Gamma_1 X \,\Gamma^\dagger_1 &=& Y, \\ 
\Gamma_1 Y \,\Gamma^\dagger_1 &=& Z, \\
\Gamma_1 Z \,\Gamma^\dagger_1 &=& X,
\end{eqnarray*}
and $\Gamma_2$ can be used to define a basis transformation to rotate Pauli observables counterclockwise, i.e., 
\begin{align*}
\label{eq:pauliop_to_pauliop2}
\Gamma_2 X \,\Gamma^\dagger_2=Z, \\ 
\Gamma_2 Y \,\Gamma^\dagger_2=X, \\
\Gamma_2 Z \,\Gamma^\dagger_2=Y.
\end{align*}

We then define a Pauli rotation $\boldsymbol{\Gamma}_\beta$ acting on $n$ qubits and identified by integer index $\beta$, with $0\le \beta\le 3^n-1$, as
\begin{equation}
\label{eq:general_gamma}
    {\boldsymbol{\Gamma}}_{\beta} = \Gamma_{\beta_1}\otimes \Gamma_{\beta_2}\otimes\text{...}\otimes \Gamma_{\beta_n},
\end{equation}
where $\beta=\overline{\beta_1 \beta_2 \text{...} \beta_n}$ is a decomposition into $n$ base-$3$ digits.

This way, we can define all the CSCOs considered in our procedure:
\begin{equation}
\label{eq:general_CSCO}
    \mathcal{C}^{\text{even}}_{\beta}={\boldsymbol{\Gamma}}_{\beta}\,\mathcal{C}^{\text{even}}_{0}\,{\boldsymbol{\Gamma}}_{\beta}^\dagger,
\end{equation}
and
\begin{equation}
\label{eq:general_CSCO2}
    \mathcal{C}^{\text{odd}}_{\beta}={\boldsymbol{\Gamma}}_{\beta}\,\mathcal{C}^{\text{odd}}_{0}\,{\boldsymbol{\Gamma}}_{\beta}^\dagger,
\end{equation}
where the basis transformation given by ${\boldsymbol{\Gamma}}_{\beta}$ acts on each element of $\mathcal{C}^{\text{even}}_{0}$ and $\mathcal{C}^{\text{odd}}_{0}$, maintaining the same ordering.
Importantly, for $n \ge 4$ all $\mathcal{C}^{\text{even}}_{\beta}$ are unique~\cite{Gatti:2021aou}, because 1) $\mathcal{C}^{\text{even}}_{\beta}(0)$ can serve as identifier (or \textit{generator}) of its CSCO, as it is the only observable that shares no Pauli-term (in the same qubit position) with any other observable $\mathcal{C}^{\text{even}}_{\beta}(i \neq 0)$ in its CSCO, and 2) it is unique among \textit{even} CSCOs as well, i.e., $\mathcal{C}^{\text{even}}_{\beta}(0)=\mathcal{C}^{\text{even}}_{\beta'}(0)\Leftrightarrow {\beta}={\beta'}$.
Similarly, all $\mathcal{C}^{\text{odd}}_{{\beta}}$ are unique as well, and there's evidently no overlap between even and odd CSCOs.

Each of the CSCOs considered has a unique basis of $2^n$ eigenstates.
We refer to the CSCOs as \textit{contexts}, and to these eigenstates as \textit{context eigenstates}.
One of the eigenstates of $\mathcal{C}^{\text{even}}_{0}$ is the $n$-qubit Greenberger-Horne-Zeilinger (GHZ) state $\ket{\phi_0,\mathcal{C}^{\text{even}}_{0}} =\ket{\text{GHZ}} \equiv \tfrac{1}{\sqrt{2}}(\ket{00\text{...}0}+\ket{11\text{...}1})$, and one of the eigenstates of $\mathcal{C}^{\text{odd}}_{0}$ is the $n$-qubit $i$-phase GHZ state $\ket{\phi_0,\mathcal{C}^{\text{odd}}_{0}}=\ket{\text{GHZ-}i} \equiv \tfrac{1}{\sqrt{2}}(\ket{00\text{...}0}+i\ket{11\text{...}1})$.
The remaining eigenstates of their bases are obtained by applying rotations on these eigenstates, specifically Pauli $Z$ phase-flip on one of their qubits (it does not matter which) and/or Pauli $X$ flips on any qubit(s), to obtain $2^n$ unique states. 
Using binary index $z=0,1$ and integer index ${\alpha}$ with $0\le {\alpha}\le 2^{n-1}-1$ to identify each context eigenstate, the eigenstates of $\mathcal{C}^{\text{even}}_{0}$ are given by
\begin{equation}
\label{eq:eigenstates_even}   \ket{\phi_{z,\alpha},\mathcal{C}^{\text{even}}_{0}}=\Big( X^{\alpha_1}\otimes X^{\alpha_2}\otimes \text{...}\otimes X^{\alpha_{n-1}}\otimes \mathbb{1}\Big) \,\, \Big(Z^{z}\otimes \mathbb{1}^{\otimes (n-1)}\Big)\ket{\text{GHZ}},
\end{equation}
and the eigenstates of $\mathcal{C}^{\text{odd}}_{0}$ are given by 
\begin{equation}
\label{eq:eigenstates_odd}
\ket{\phi_{z,\alpha},\mathcal{C}^{\text{odd}}_{0}}=\Big( X^{\alpha_1}\otimes X^{\alpha_2}\otimes \text{...}\otimes X^{\alpha_{n-1}}\otimes \mathbb{1}\Big) \,\, \Big(Z^{z}\otimes \mathbb{1}^{\otimes (n-1)}\Big)\ket{\text{GHZ-}i},
\end{equation}
where ${\alpha}=\overline{\alpha_1 \, \alpha_2 \, \text{...} \, \alpha_{n-1}}$ is a decomposition into $n-1$ base-$2$ digits.
This way, each value of $\alpha_i$ is either $0$ or $1$ and indicates whether the Pauli flip $X$ in qubit $i$ is performed or not (recall that $X^0=\mathbb{1}$ and $X^1=X$).
Similarly, $z$ indicates  whether the Pauli phase-flip $Z$ in qubit $1$ is performed or not ($Z^0=\mathbb{1}$ and $Z^1=Z$).

To obtain the context eigenstates of any context $\mathcal{C}^{\text{even}}_{{\beta}}$ or $\mathcal{C}^{\text{odd}}_{{\beta}}$, we simply apply the Pauli rotation ${\boldsymbol{\Gamma}}_{{\beta}}$ on the eigenstates of $\mathcal{C}^{\text{even}}_{0}$ or on those of $\mathcal{C}^{\text{odd}}_{0}$, respectively, obtaining
\begin{equation}
\label{eq:all_eigenstates_even}
\ket{\phi_{z,\alpha},\mathcal{C}^{\text{even}}_{\beta}}={\boldsymbol{\Gamma}}_{{\beta}}\, \, \Big( X^{\alpha_1}\otimes X^{\alpha_2}\otimes \text{...}\otimes X^{\alpha_{n-1}}\otimes \mathbb{1}\Big) \,\, \Big(Z^{z}\otimes \mathbb{1}^{\otimes (n-1)}\Big)\ket{\text{GHZ}},
\end{equation}
and
\begin{equation}
\label{eq:all_eigenstates_odd}
\ket{\phi_{z,\alpha},\mathcal{C}^{\text{odd}}_{\beta}}={\boldsymbol{\Gamma}}_{{\beta}}\, \, \Big( X^{\alpha_1}\otimes X^{\alpha_2}\otimes \text{...}\otimes X^{\alpha_{n-1}}\otimes \mathbb{1}\Big) \,\, \Big(Z^{z}\otimes \mathbb{1}^{\otimes (n-1)}\Big)\ket{\text{GHZ-}i}.
\end{equation}

Additionally, we can merge both expressions by writing $\mathcal{C}^{\text{0}}_{\beta}$ instead of $\mathcal{C}^{\text{even}}_{\beta}$ and $\mathcal{C}^{\text{1}}_{\beta}$ instead of $\mathcal{C}^{\text{odd}}_{\beta}$, and introducing one more binary index $s=0,1$ which will indicate whether a Phase gate $S$ on qubit $1$ is performed or not, allowing to obtain $\ket{\text{GHZ-}i}$ from $\ket{\text{GHZ}}$:
\begin{equation}
\label{eq:all_eigenstates_evenodd}
\ket{\phi_{z,\alpha},\mathcal{C}^{s}_{\beta}}={\boldsymbol{\Gamma}}_{{\beta}}\, \, \Big( X^{\alpha_1}\otimes X^{\alpha_2}\otimes \text{...}\otimes X^{\alpha_{n-1}}\otimes \mathbb{1}\Big) \,\, \Big(Z^{z}\otimes \mathbb{1}^{\otimes (n-1)}\Big) \,\, \Big(S^{s}\otimes \mathbb{1}^{\otimes (n-1)}\Big)\ket{\text{GHZ}}.
\end{equation}

\begin{figure*}[ht!]
\includegraphics[width=0.8\textwidth]{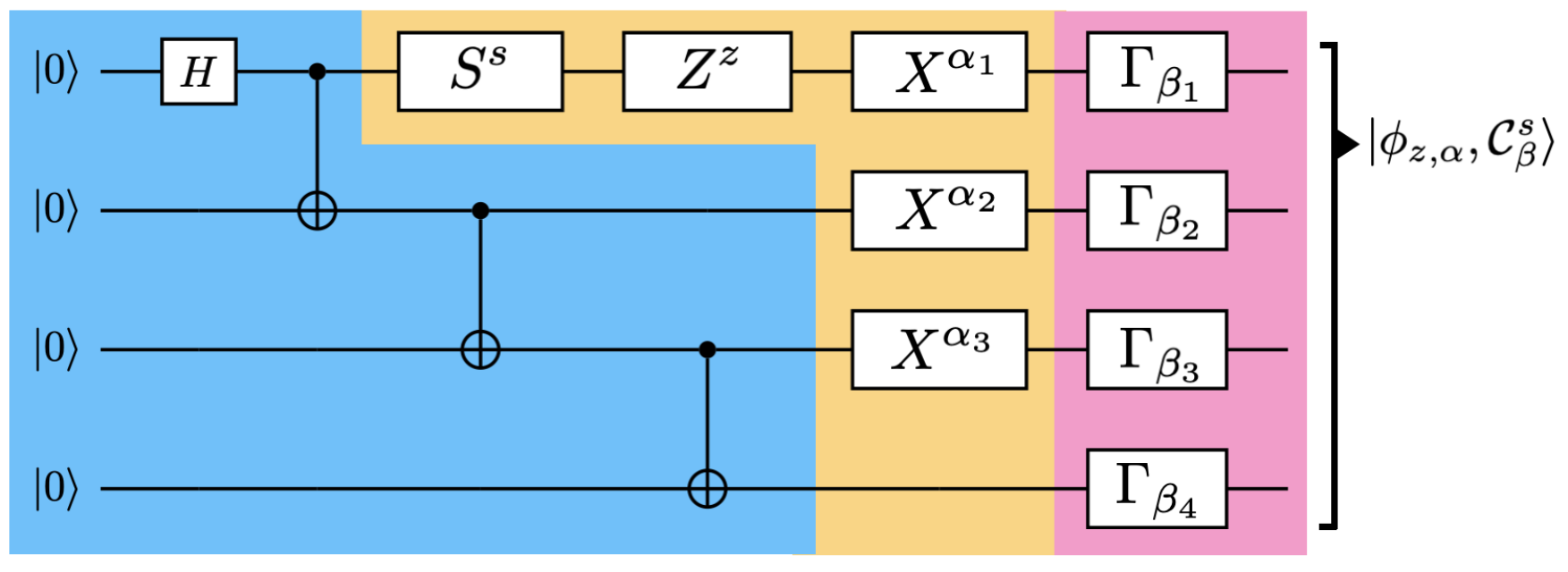}
\caption{\textbf{\textit{Context eigenstate preparation.}} 
Quantum circuit to prepare the $4$-qubit context eigenstate $\ket{\phi_{z,\alpha},\mathcal{C}^{s}_{\beta}}$.
The blue section of the circuit trivially prepares a $4$-qubit GHZ state $\ket{GHZ}=\tfrac{1}{\sqrt{2}}(\ket{0000}+\ket{1111})$, the yellow section performs local rotations based on binary parameters $s$, $z$, $\alpha_1$, $\alpha_2$ and $\alpha_3$, and the pink section performs local rotations based on ternary parameters $\beta_1$, $\beta_2$, $\beta_3$ and $\beta_4$.
Note that this didactic circuit prepares the GHZ state with depth $O(n)$ ($n-1$ layers of two-qubit-gates), but GHZ states can be prepared with lower depth if the qubits have more connectivity than what is used here, as will be seen in the next section. 
}
\label{fig:circuit_preparation}
\end{figure*}

We can note that this expression allows us to easily prepare any $n$-qubit context eigenstate $\ket{\phi_{z,\alpha},\mathcal{C}^{s}_{\beta}}$ in a quantum circuit, provided that we prepare a GHZ state first.
An example is shown in Suppl.~Fig.~\ref{fig:circuit_preparation} for $n=4$ qubits, where we can note that only the GHZ state preparation requires the use of two-qubit gates.
The rest of the circuit consists in local operations which depend on the value of the indexes $s$, $z$, $\alpha$ and $\beta$.
The indexes $s$ and $\beta$ define the context $\mathcal{C}^{s}_{\beta}$ where the eigenstate is prepared (from $2\times 3^n$ options), and $z$ and $\alpha$ define the eigenstate $\phi_{z,\alpha}$ itself (from $2^n$ options).
In the Genetic Algorithm training in this work, we employ tuples of these indexes to identify each context eigenstate.

%% file: app-eff-parity.tex
\section{Efficient Calculation of Parity}
\label{appendix:fingerprint}

With the notation and definitions of~\ref{appendix:PO_eigenstates}, we are able to analytically calculate the parity of any context eigenstate.
By definition, the expected outcome of measuring an $n$-qubit quantum state with observable $\mathcal{O} \in \{X,Y,Z\}^{\otimes n}$ lies between $-1$ and $+1$ (inclusive), since those are the eigenvalues of $\mathcal{O}$.
Notably, if a context eigenstate is measured, the expected outcome only assumes one of three possible values $\bra{\phi_{z,\alpha},\mathcal{C}^{s}_{\beta}} \mathcal{O} \ket{\phi_{z,\alpha},\mathcal{C}^{s}_{\beta}}=+1,-1,0$, the latter when the eigenstate's context $\mathcal{C}^{s}_{\beta}$ does not include observable $\mathcal{O}$.
To take advantage of mod-$2$ notation, we map these values to binary via the function ${\rm{bit}}(\lambda)=(1-\lambda)/2$.
This way, we define the parity of measuring an arbitrary state $\ket{\psi}$ in $\mathcal{O} \in \{X,Y,Z\}^{\otimes n}$ as
\begin{equation}
\mathcal{P}\Big(\ket{\psi},\mathcal{O}\Big)=\rm{bit}\Big(\bra{\psi}\mathcal{O} \ket{\psi}\Big),
\end{equation}
such that a context eigenstate yields one of three parities $\mathcal{P}(\ket{\phi_{z,\alpha},\mathcal{C}^{s}_{\beta}},\mathcal{O})=0\,,\,1\,,\,0.5$, which we call \textit{even} parity, \textit{odd} parity and \textit{unbiased} parity, respectively.

Conveniently, when the measurement is performed in $Z^{\otimes n}$, canonical-basis states with even number of $1$s (like $\ket{01010101}$) yield even parity, and so do linear combinations of them. 
Similarly, canonical-basis states with odd number of $1$s (like $\ket{01010111}$) yield odd parity, and so do linear combinations of them.

Briefly note that the set of parity observables $\{X,Y,Z\}^{\otimes n}$ can be indexed and ordered using $\boldsymbol{\Gamma}_{\beta'}$ rotations on $Z^{\otimes n}$, i.e.,
\begin{equation}
\mathcal{O}_{\beta'}=\boldsymbol{\Gamma}_{\beta'}\, Z^{\otimes n} \,\boldsymbol{\Gamma}^\dagger_{\beta'}.
\end{equation}
This way, the $\mathcal{O}_{\beta'}$ parity of context eigenstate $\ket{\phi_{z,\alpha},\mathcal{C}^{s}_{\beta}}$ is given by
\begin{equation}
\label{eq:parity_def}
\mathcal{P}\Big(\ket{\phi_{z,\alpha},\mathcal{C}^{s}_{\beta}},\mathcal{O}_{\beta'}\Big)=\rm{bit}\Big(\bra{\phi_{z,\alpha},\mathcal{C}^{s}_{\beta}}\boldsymbol{\Gamma}_{\beta'}\, Z^{\otimes n} \,\boldsymbol{\Gamma}^\dagger_{\beta'} \ket{\phi_{z,\alpha},\mathcal{C}^{s}_{\beta}}\Big),
\end{equation}
which is equivalent to finding the $Z^{\otimes n}$ parity of state $\boldsymbol{\Gamma}^\dagger_{\beta'} \ket{\phi_{z,\alpha},\mathcal{C}^{s}_{\beta}}$:
\begin{equation}
\label{eq:parity_def2}
\mathcal{P}\Big(\ket{\phi_{z,\alpha},\mathcal{C}^{s}_{\beta}},\mathcal{O}_{\beta'}\Big)=\mathcal{P}\Big(\boldsymbol{\Gamma}^\dagger_{\beta'}\ket{\phi_{z,\alpha},\mathcal{C}^{s}_{\beta}},Z^{\otimes n}\Big).
\end{equation}

We can compute the parity by considering three cases for the relationship between $\beta$ and $\beta'$, as well as their base-$3$ decomposition $\beta=\overline{\beta_1 \, \beta_2 \, \text{...} \,\beta_n}$ and $\beta'=\overline{\beta_1' \, \beta_2' \, \text{...} \,\beta_n'}$:

\textbf{Case 1:} $\beta=\beta'$

In this scenario, the $\boldsymbol{\Gamma}^\dagger_{\beta'}$ and $\boldsymbol{\Gamma}_\beta$ rotations cancel out such that the $Z^{\otimes n}$ parity of $\boldsymbol{\Gamma}^\dagger_{\beta} \ket{\phi_{z,\alpha},\mathcal{C}^{s}_{\beta}}$ is directly given by the total number of $X$ gates that are $\textit{turned ON}$. Consequently,
\begin{equation}
\label{eq:parity_analytic_p1}
\mathcal{P}\Big(\ket{\phi_{z,\alpha},\mathcal{C}^{s}_{\beta}},\mathcal{O}_{\beta'}\Big)=\bigoplus_{i=1}^{n-1} \,{\alpha_i},
\end{equation}
where $\oplus$ denotes sum modulo $2$.

\textbf{Case 2:} $\exists \,i,j \mid \beta_i=\beta'_i\,,\,\beta_j\neq\beta'_j$

In this scenario, $\beta$ shares at least one ternary digit with $\beta'$ and has at least one different ternary digit.
Since each observable in context $\mathcal{C}^{s}_{\beta}$ either shares no term with the \textit{generator} $\mathcal{C}^{s}_{\beta}(0)=\mathcal{O}_\beta$ or shares them all (in which case it is the actual generator), then $\mathcal{O}_{\beta'}$ is not in $\mathcal{C}^{s}_{\beta}$, meaning that the parity is unbiased:
\begin{equation}
\label{eq:parity_analytic_p2}
\mathcal{P}\Big(\ket{\phi_{z,\alpha},\mathcal{C}^{s}_{\beta}},\mathcal{O}_{\beta'}\Big)=0.5.
\end{equation}

\textbf{Case 3:} $\beta_i \neq \beta'_i \,\,\forall i$

In this scenario $\mathcal{O}_{\beta'}$ is either in context $\mathcal{C}^{s=0}_{\beta}$ or $\mathcal{C}^{s=1}_{\beta}$, which means that the parity will be unbiased for some value of $s$ and not unbiased for the other.
To express the parity, we define  vectors of length $n$ based on the binary and ternary decomposition of the indexes $\boldsymbol{\alpha}=(\alpha_1,\alpha_2,\text{...},\alpha_{n-1},0)$, $\boldsymbol{\beta}=(\beta_1,\beta_2,\text{...},\beta_{n})$ and $\boldsymbol{\beta'}=(\beta'_1,\beta'_2,\text{...},\beta'_{n})$.
Note that $\boldsymbol{\alpha}$ has been zero-padded to have length $n$.
Using~\cref{eq:parity_def2}, and counting the number of $1$s in the canonical-basis expression for $\boldsymbol{\Gamma}^\dagger_{\beta'}\ket{\phi_{z,\alpha},\mathcal{C}^{s}_{\beta}}$, we obtain
\begin{equation}
\label{eq:parity_analytic_p3}
\mathcal{P}\Big(\ket{\phi_{z,\alpha},\mathcal{C}^{s}_{\beta}},\mathcal{O}_{\beta'}\Big)=z \oplus \big( \text{sum}(\mathbf{J})+s\big)/2 \oplus \mathbf{J}.\boldsymbol{\alpha},
\end{equation}
where $\mathbf{J}.\boldsymbol{\alpha}$ is a dot product, and we define $\mathbf{J}=\vec{2}-(\boldsymbol{\beta}\ominus_3 \boldsymbol{\beta'})$, where $\ominus_3$ denotes subtraction modulo $3$ and $\vec{2}=(2,2,\text{...},2)$ is a vector of length $n$.
We can note that depending on the value of $s$, the term $( \text{sum}(\mathbf{J})+s\big)/2$ is an integer or a half-integer.
This allows us to obtain both the $\mathcal{O}_{\beta'} \in \mathcal{C}^{s}_{\beta}$ case, where the parity is $0$ (even) or $1$ (odd), and the
$\mathcal{O}_{\beta'} \notin \mathcal{C}^{s}_{\beta}$ case, where the parity is $0.5$ (unbiased).

Storing the parity values $\mathcal{P}\left(\ket{\phi_{z,\alpha},\mathcal{C}^{s}_{\beta}},\mathcal{O}_{\beta'}\right)$ for $s=0,1$, for all $z$,${\alpha}$ and for all $\mathbf{J}$ satisfying $\beta_i \neq \beta'_i \,\,\forall i$ requires $2\times 2^n \times 2^n$ bits.
We refer to this collection of bits as the even ($s=0$) and odd ($s=1$) \textit{fingerprints}. 
Precalculating and storing them allows us to efficiently calculate any context eigenstate parity. We do this in our $8$-qubit implementation.

%% file: app-context-measurement.tex
\section{Context measurement}
\label{app:context_measurement}

It is trivial to measure a state with observable $\mathcal{O}_0=Z^{\otimes n}$, by performing regular (canonical) measurements on each qubit and counting the number of $1$s.
Obtaining an even (odd) number of $1$s is equivalent to the outcome $+1$ ($-1$) of observable $\mathcal{O}_0$.

Similarly, it is also trivial to measure a state with observable $\mathcal{O}_{\beta'}=\boldsymbol{\Gamma}_{\beta'} Z^{\otimes n} \boldsymbol{\Gamma}^\dagger_{\beta'}$, rotating it via $\boldsymbol{\Gamma}^\dagger_{\beta'}$ and then performing regular (canonical) measurements on each qubit, counting the number of $1$s.
Obtaining an even (odd) number of $1$s is equivalent to the outcome $+1$ ($-1$) of observable $\mathcal{O}_{\beta'}$.

However, both of these measurements fully collapse an $n$-qubit quantum state and yield a single bit ($+1$ or $-1$), whereas we would expect to obtain one in $2^n$ possible outcomes ($n$ bits) when measuring in an $n$-qubit basis.
Specifically, it should be possible to measure in a complete (sub)set of commuting observables $\mathcal{C}^{s'}_{\beta'}$ and obtain the parity for each observable in the set, by collapsing into one of the $2^n$ context eigenstates $\{\ket{\phi_{z,\alpha},\mathcal{C}^{s'}_{\beta'}}\}_{z,\alpha}$.
Note that we are adding primes to $s$ and $\beta$ in the measured context $\mathcal{C}^{s'}_{\beta'}$, so as to distinguish from the parameters used in the state-generation.

\begin{figure*}[ht!]
\includegraphics[width=0.75\textwidth]{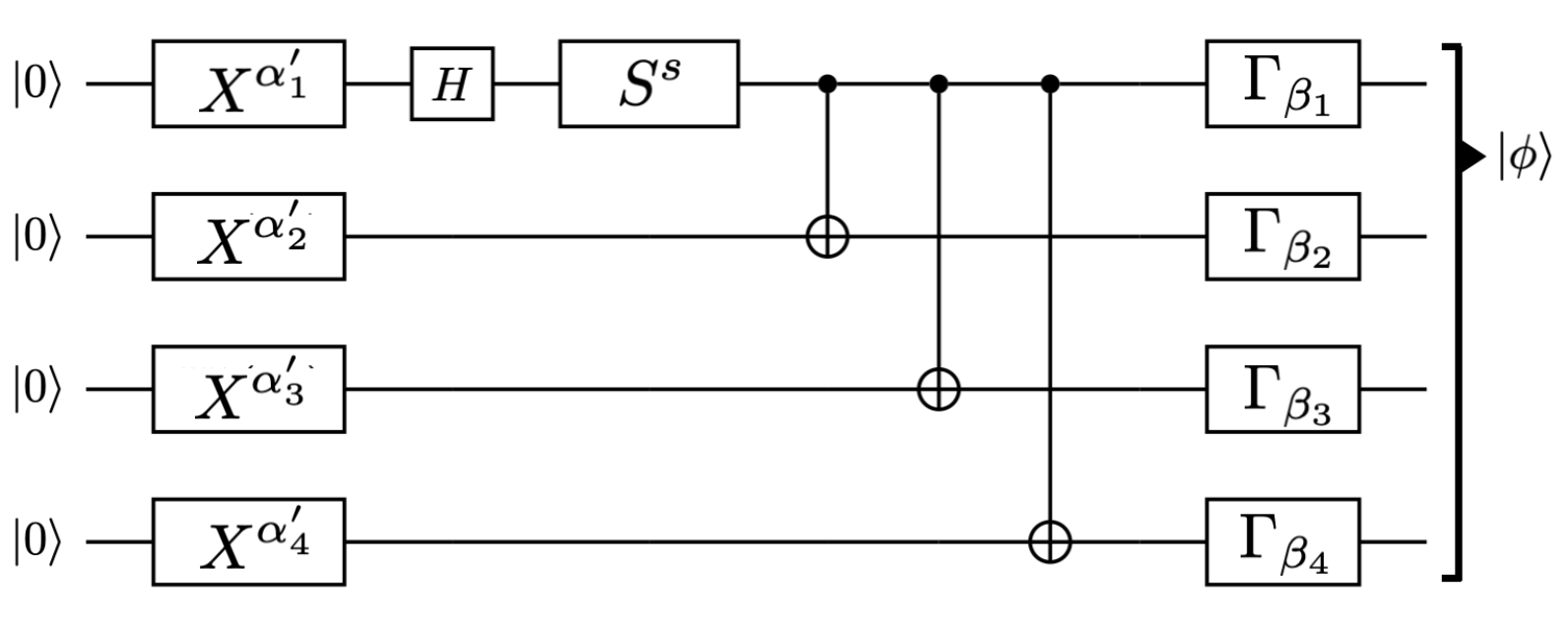}
\caption{\textbf{\textit{Alternative context eigenstate preparation.}} 
Alternative quantum circuit to prepare a context eigenstate denoted by $\ket{\phi}$, belonging to the same class as $\ket{\phi_{z,\alpha},\mathcal{C}^s_\beta}$ but parametrized in a different manner. As before, the circuit is showcased for $n=4$ qubits, but is generalizable to any $n$.
The original circuit has been modified so that the first gates applied are Pauli $X$ gates which decide which of $2^n$ eigenstates to prepare.
This allows to use the inverse of this circuit to perform a full-context measurement, by mapping context eigenstates to canonical-basis elements.
The two-qubit gates employed in this circuit are harder to implement with low depth than those in Suppl.~Fig.~\ref{fig:circuit_preparation}, which is why this circuit is not used for state generation in our implementation.
}
\label{fig:circuit_preparation2}
\end{figure*}

\begin{figure*}[ht!]
\includegraphics[width=0.75\textwidth]{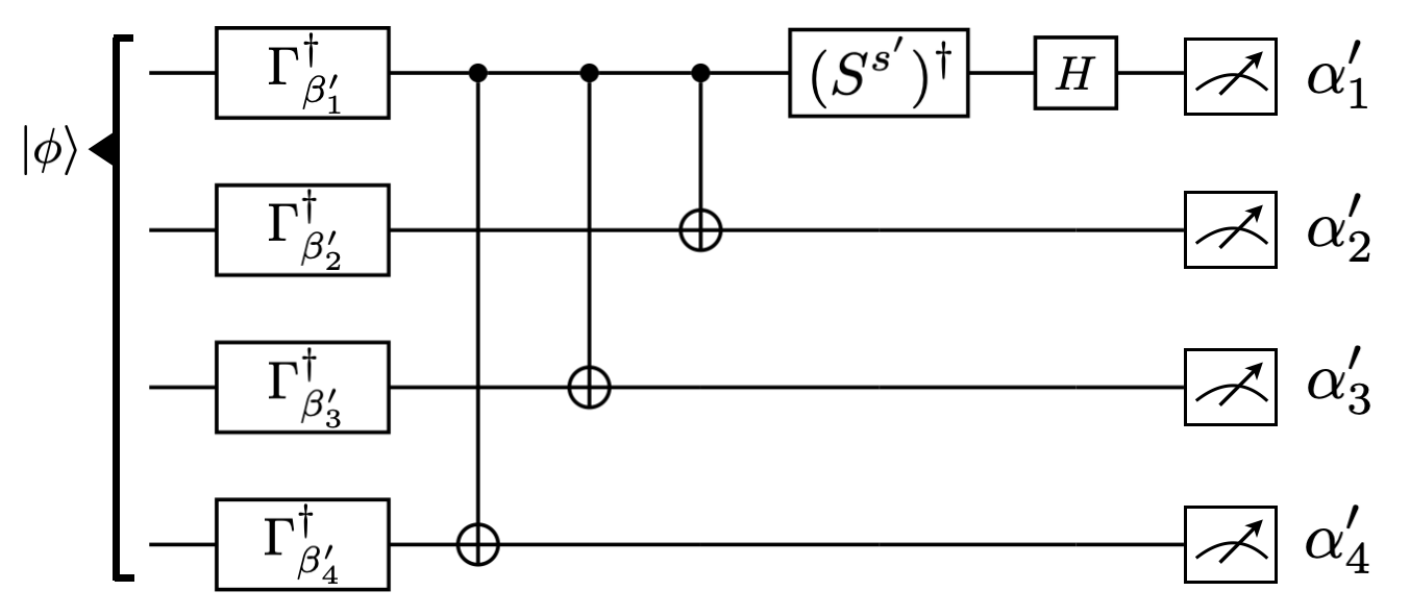}
\caption{\textbf{\textit{Context measurement.}} 
Quantum circuit to measure an arbitrary input state $\ket{\phi}$ in context $\mathcal{C}^{s'}_{\beta'}$. The circuit is showcased for $n=4$ qubits, but is generalizable to an arbitrary number of qubits.
To achieve the desired measurement, this circuit maps eigenstates from context $\mathcal{C}^{s'}_{\beta'}$ into canonical-basis elements, and then measures in that basis, obtaining outcomes $\{\alpha'_1,\alpha'_2,\alpha'_3,\alpha'_4\}$. 
}
\label{fig:context_measurement}
\end{figure*}

A circuit similar to the inverse of Suppl.~Fig.~\ref{fig:circuit_preparation} could be used to map context eigenstates $\{\ket{\phi_{z,\alpha},\mathcal{C}^{s'}_{\beta'}}\}_{z,\alpha}$ to the canonical basis.
For this end, consider the alternative state-generation circuit depicted in  Suppl.~Fig.~\ref{fig:circuit_preparation2}, which produces the same set of eigenstates $\{\ket{\phi_{z,\alpha},\mathcal{C}^{s}_{\beta}}\}_{z,\alpha}$ in~\cref{eq:all_eigenstates_evenodd}, but with slightly different parameters $\{\ket{\phi_{\alpha'},\mathcal{C}^{s}_{\beta}}\}_{\alpha'}$.
We employ the inverse of this circuit to construct a context-measurement circuit, and showcase it for $n=4$ qubits in Suppl.~Fig.~\ref{fig:context_measurement}.

In its general form, this circuit produces the following evolution on an arbitrary $n$-qubit input state $\ket{\psi}$
\begin{equation}
\label{eq:context_eigenstate_to_canonical}
\mathcal{M}(s',\beta')\ket{\psi}=\Big(H\otimes \mathbb{1}^{\otimes (n-1)}\Big)\,\, \Big({S^{s'}}^{\dagger}\otimes \mathbb{1}^{\otimes (n-1)}\Big)\,\,\Bigg(\prod_{i=2}^{n} \text{CX}(1,i)\Bigg)\,\,\Big(\boldsymbol{\Gamma}_{\beta'}^\dagger\Big)\ket{\psi},
\end{equation}
after which it is measured in the canonical basis. Here, we have defined $\text{CX}(i,j)$ as the two-qubit Control-NOT gate from qubit $i$ (control) to qubit $j$ (target) and the identity $\mathbb{1}$ on all other qubits, so that each $\text{CX}(i,j)$ acts on the $n$-qubit phase space.
The canonical measurement outputs binary values $\{\alpha'_1,\alpha'_2,\text{...},\alpha'_n\}$, which we can map back to $z,\alpha$ parameters in the context eigenstate notation $\{\ket{\phi_{z,\alpha},\mathcal{C}^{s'}_{\beta'}}\}_{z,\alpha}$ using the following rules:
\begin{align*}
\label{eq:alphaprime_to_alpha}
\alpha_1=\alpha'_n \\ 
\alpha_j=\alpha'_j \oplus \alpha'_n \,\,\,\,\,,\,\,\,\,\, 2\le j \le n-1 \\
z=\alpha'_1 \oplus (s' \alpha'_n).
\end{align*}

This way, an $n$-qubit quantum state $\ket{\psi}$ can be measured in context $\mathcal{C}^{s'}_{\beta'}$ by evolving it with $\mathcal{M}(s',\beta')$ and measuring in the canonical basis.
The canonical-measurement outcomes $\{\alpha'_1,\alpha'_2,\text{...},\alpha'_n\}$ are mapped to $z,\alpha$ with the rules above, and the result of the overall measurement is that the system $\ket{\psi}$ has collapsed to state $\ket{\phi_{z,\alpha},\mathcal{C}^{s'}_{\beta'}}$. Since the actual state of the system is now available to us from the measurement outcome, we can use~\cref{eq:parity_analytic_p1} and \eqref{eq:parity_analytic_p3} to analytically determine the parity for each observable $\mathcal{O}_{\beta''}$ in context $\mathcal{C}^{s'}_{\beta'}$, effectively measuring $\ket{\psi}$ in those observables.

%% file: app-circuit-optimization.tex
\section{Circuit Depth Optimization on IBM Cairo}
\label{app:depth_optim}

\begin{figure*}[ht!]
\includegraphics[width=0.7\textwidth]{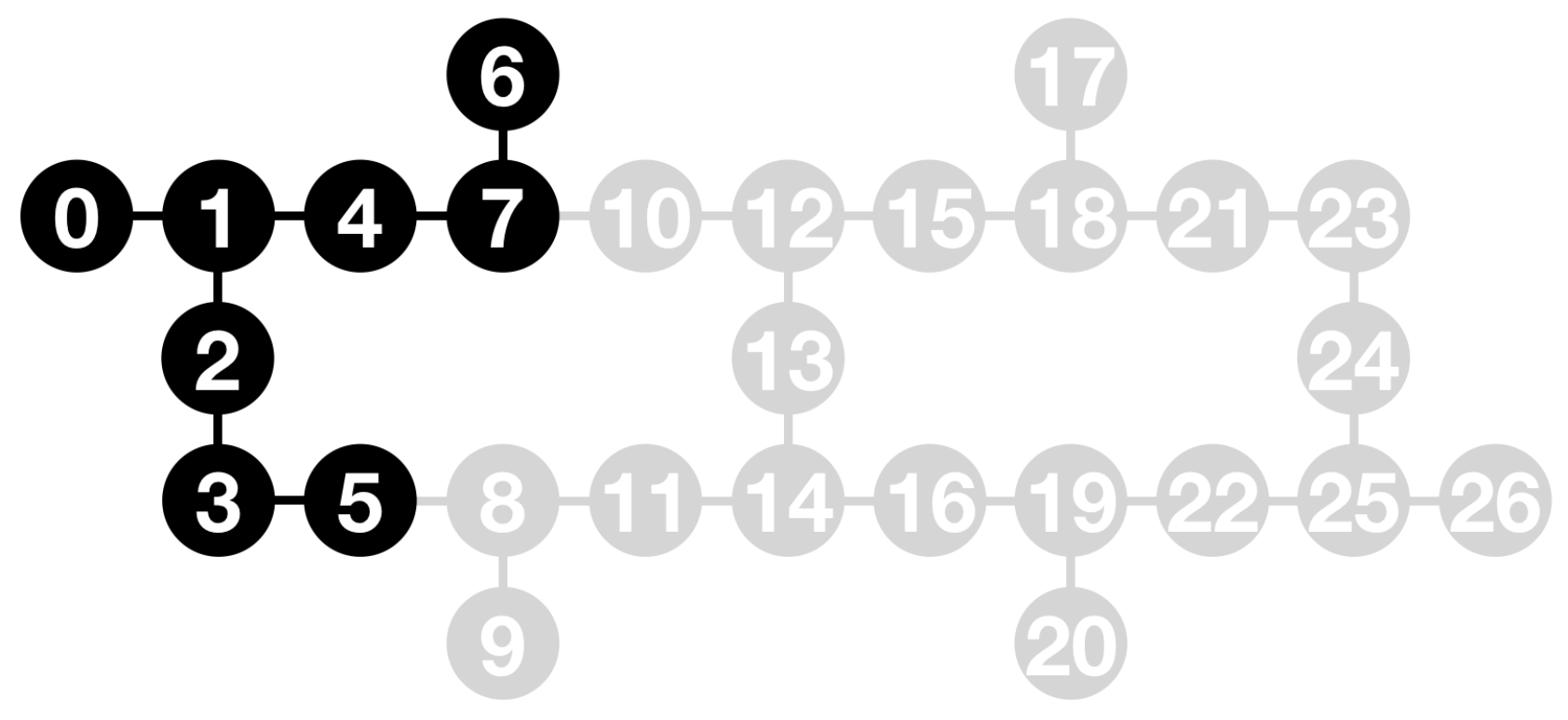}
\caption{\textbf{\textit{IBM-Qiskit Cairo connectivity.}} 
Map of the two-qubit gate connectivity in the Cairo real backend of IBM-Qiskit.
Each circle represents a qubit with its corresponding index, and each line represents the possibility of directly performing two-qubit gates between the connected qubits.
In our work, we employed the qubits (and connectivity lines) in black, selecting them due to appropriate connectivity for our scheme, and lower error rates in a GHZ preparation respect to other qubits at the time.
}
\label{fig:cairo_connectivity}
\end{figure*}

In this work, we have employed the state generation and measurement circuits in Suppl.~Figs.~\ref{fig:circuit_preparation} and \ref{fig:context_measurement}, generalizing them to systems of $n=8$ qubits, as explicitly laid out in Eqs.~\eqref{eq:all_eigenstates_evenodd} and \eqref{eq:context_eigenstate_to_canonical} (followed by a canonical measurement).
It is desirable in their actual implementation to minimize the number of \textit{layers} with two-qubit gates, also known as \textit{circuit depth}, so as to minimize decoherence.
In our state generation circuit, this means finding a low-depth implementation to prepare a GHZ state, and in our measurement circuit this means finding a low-depth implementation of $\prod_{i=2}^n \text{CX}(1,i)$, that is, a sequence of Control-NOTs from qubit $1$ (control) to all other qubits (target).
The implementations need to take into account the architecture used as well.
We have employed the IBM-Qiskit Cairo backend, with two-qubit-gate connectivity shown in Suppl.~Fig.~\ref{fig:cairo_connectivity}.

Regarding the state generation circuit, note that a $k$-qubit GHZ state can be turned into a $2 k$-qubit GHZ state in a single two-qubit layer if the architecture has sufficient connectivity.
This is done by applying CNOTs from each of the $k$ qubits (control) conforming the GHZ state into $k$ qubits (target) initialized in $\ket{0}$.
Thus, a GHZ state can be constructed exponentially fast and an $8$-qubit GHZ state can be constructed with as few as $3$ two-qubit-gate layers.
With Cairo's connectivity, we are able to construct it with $4$ layers, in the following manner:
\begin{equation}
\label{eq:cairo_GHZ}
\ket{\text{GHZ}}=\Big(\text{CX}(7,6)\Big)\,\,\Big(\text{CX}(3,5)\,\text{CX}(4,7)\,\text{CX}(1,0)\Big)\,\,\Big(\text{CX}(2,3)\,\text{CX}(1,4)\Big)\,\,\Big(\text{CX}(1,2)\Big)\,\,H_1\ket{00000000},
\end{equation}
where we explicitly indicate each two-qubit-gate layer with big parentheses, and where $H_1$ denotes Hadamard $H$ on qubit index $1$ and identity $\mathbb{1}$ on all other qubits.
The actual order of the qubits in $\ket{00000000}$ is irrelevant, as the gates have been specified in terms of their indexes.

Regarding the measurement circuit, an operation that is equivalent to Control-NOTs from one qubit to all others cannot grow exponentially fast in the number of two-qubit-gate layers.
Instead $k$ qubits with linear connectivity can do it in $2 k-3$ layers such that the common control qubit is operated on just once (the operation order is read from right to left):
\begin{equation}
\label{eq:common_control_arm}
\prod_{i=2}^k \text{CX}(1,i)=\Big(\text{CX}(k-1,k)\Big)\,\Big(\text{CX}(k-2,k-1)\Big)\,\text{...}\,\Big(\text{CX}(2,3)\Big)\,\Big(\text{CX}(1,2)\Big)\,\Big(\text{CX}(2,3)\Big)\,\text{...}\,\Big(\text{CX}(k-2,k-1)\Big)\,\Big(\text{CX}(k-1,k)\Big),
\end{equation}
where we explicitly indicate each two-qubit-gate layer with big parentheses.

This allows to construct an operation in Cairo equivalent to CNOTs from qubit index $1$ (control) to all others (target) by considering three groups of linearly-connected qubits in the architecture: $\{1,2,3,5\}$,$\{1,4,7,6\}$ and $\{1,0\}$. The operations in~\cref{eq:common_control_arm} allow to use $5$, $5$ and $1$ layers respectively for each group, requiring to operate on qubit index $1$ in the $3^{\text{rd}}$, $3^{\text{rd}}$ and $1^{\text{st}}$ (and only) layer respectively.
The layer in which qubit $1$ is operated on is important because that is the only overlap between the groups, and all other two-qubit gates can be done in parallel.
This way, the operations for group $\{1,2,3,5\}$ and group $\{1,0\}$ can start at the same time, while group $\{1,4,7,6\}$ is delayed one layer, such that its operation over qubit $1$ happens at the $4^{\text{th}}$ layer and does not conflict with group $\{1,2,3,5\}$.
This allows to perform $\prod_{i\neq 1} \text{CX}(1,i)$ over $n=8$ qubits in Cairo using only $6$ two-qubit-gate layers, as shown in the following (the operation order is read from right to left):

\begin{multline}
\label{eq:common_control_full}
\prod_{i\neq 1} \text{CX}(1,i)=\Big(\text{CX}(7,6)\Big)\,\, \Big(\text{CX}(3,5)\,\text{CX}(4,7)\Big)\,\, \Big(\text{CX}(2,3)\,\text{CX}(1,4)\Big)\\\Big(\text{CX}(1,2)\,\text{CX}(4,7)\Big)\,\,\Big(\text{CX}(2,3)\,\text{CX}(7,6)\Big)\,\,\Big(\text{CX}(1,0)\,\text{CX}(3,5)\Big),
\end{multline}
where we explicitly indicate each two-qubit-gate layer with big parentheses.

This way, our $8$-qubit state generation circuit is constructed in a real backend with a depth of $4$ two-qubit-gate layers, and the corresponding measurement circuit is constructed with a depth of $6$ two-qubit-gate layers, taking advantage of Cairo's connectivity surrounding qubit $1$.
This relatively low depth of $10$ layers allows us to obtain the very high experimental fidelity that we report in our conclusions.

%% file: app-context-eigenstates-in-code.tex
\section{Context eigenstates in code}
\label{appendix:context_eigenstates_in_code}

In this work, the data of each simulated IceCube event is encoded into a selection of $N$ context eigenstates of $n=8$ qubits, for variable values of $N$. We denote this selection by 
\begin{equation}
\Big\{\ket{\phi}\!\Big\}\equiv \bigg\{\ket{\phi_{z\mbox{\tiny $(i)$},\alpha\mbox{\tiny $(i)$}},\mathcal{C}^{s\mbox{\tiny $(i)$}}_{\beta\mbox{\tiny $(i)$}}} \,\,\bigg| \,\,i=1,2,\text{...},N\bigg\}\text{.}
\end{equation}

To decide which context eigenstate selection to employ for the encoding, it is first necessary to describe the context eigenstates $\ket{\phi_{z\mbox{\tiny $(i)$},\alpha\mbox{\tiny $(i)$}},\mathcal{C}^{s\mbox{\tiny $(i)$}}_{\beta\mbox{\tiny $(i)$}}}$ in a form that can be incorporated into regular code. The obvious solution is writing down a tuple of their parameters. We do it in the following manner

\begin{equation}
\ket{\phi_{z\mbox{\tiny $(i)$},\alpha\mbox{\tiny $(i)$}},\mathcal{C}^{s\mbox{\tiny $(i)$}}_{\beta\mbox{\tiny $(i)$}}}\Rightarrow \bigg(\mathcal{C}^{(i)}_\text{ID}\,,\,\text{str}\Big(s\mbox{\small $(i)$}\Big)+\text{str}\Big(z\mbox{\small $(i)$}\Big)+\text{str}\Big(\text{binary}\big(\alpha\mbox{\small $(i)$}\big)\Big)\,,\,\text{str}\Big(\text{ternary}\big(\beta\mbox{\small $(i)$}\big)\Big)\bigg)\text{,}
\end{equation}

\noindent where $s{(i)}$ (bit), $z{(i)}$ (bit) and $\alpha{(i)}$ in binary form ($n-1$ bitstring) are joined into a larger string of $0$s and $1$s, followed by $\beta{(i)}$ in ternary form ($n$ trit-string). Additionally,  we include a redundant parameter $\mathcal{C}^{(i)}_\text{ID}$ built from $\beta{(i)}$ and $s{(i)}$ to uniquely identify every context $\mathcal{C}^{s\mbox{\tiny $(i)$}}_{\beta\mbox{\tiny $(i)$}}$,

\begin{equation}
\mathcal{C}^{(i)}_\text{ID}=\beta{(i)}+s{(i)} \times 3^n\text{.}
\label{eq:context_id}
\end{equation}

\noindent where we should note that $\beta{(i)}$ is an integer between $0$ and $3^n-1$, such that $\mathcal{C}_\text{ID}$ is an integer between $0$ and $2 \times 3^n -1$.

Additionally, recall that the context generation and measurement circuits make use of gamma rotations $\boldsymbol{\Gamma}_\beta$, defined in Eqs.~\eqref{eq:pauli_to_pauli} and \eqref{eq:general_gamma}. Specifically, the last step of a generation circuit is a $\boldsymbol{\Gamma}_\beta$ rotation, and the first step of a measurement circuit is a $\boldsymbol{\Gamma}^\dagger_{\beta'}$ rotation. These rotations are evolutions acting on $n$-qubit Hilbert space, but it is possible to simplify their product by considering that it satisfies

\begin{equation}
\boldsymbol{\Gamma}_{\beta'} \, \boldsymbol{\Gamma}_{\beta}=\boldsymbol{\Gamma}_{\beta''}\,\,\Big|\,\,\beta''_i=\beta_i +\beta'_i\, (\text{mod }3)\,\,\,\forall i\in\{1,2,\text{...},n\}
\end{equation}

\noindent and

\begin{equation}
\boldsymbol{\Gamma}^\dagger_{\beta'} \, \boldsymbol{\Gamma}_{\beta}=\boldsymbol{\Gamma}_{\beta''}\,\,\Big|\,\,\beta''_i=\beta_i -\beta'_i \,(\text{mod }3)\,\,\,\forall i\in\{1,2,\text{...},n\}
\end{equation}

\noindent with $\beta=\overline{\beta_1 \beta_2 \text{...} \beta_n}$, $\beta'=\overline{\beta'_1 \beta'_2 \text{...} \beta'_n}$ and $\beta''=\overline{\beta''_1 \beta''_2 \text{...} \beta''_n}$ representing decompositions into $n$ base-$3$ digits. 

Considering these properties, in our code we define a class called \textit{OpString} for trit-strings (e.g. $\text{`}0211\text{'}$) where we incorporate addition-mod-$3$,
\begin{equation}
\text{\textbf{e.g.}}\,\,\text{`}0211\text{'}+\text{`}2222\text{'}=\text{`}2100\text{'}\text{,}
\end{equation}
\noindent subtraction-mod-$3$,
\begin{equation}
\text{\textbf{e.g.}}\,\,\text{`}0211\text{'}-\text{`}2222\text{'}=\text{`}1022\text{'}\text{,}
\end{equation}
\noindent and mod-$3$-complement
\begin{equation}
\text{\textbf{e.g.}}\,\,\text{complement}(\text{`}0211\text{'})=\text{`}0122\text{'}\text{,}
\end{equation} 
\noindent such that these operations are applied at the level of each digit. To make an element of this class from a trit-string, we apply $\text{OpString}(\cdot)$ on a string representation of it, e.g.

\begin{python}
ops_1=OpString('0211')
ops_2=OpString('2222')
ops_3=ops_1+ops_2#ops_3 is the same as OpString('2100')
\end{python}

For reference, we provide below the \textit{OpString} class definition:

\begin{python}
class OpString(str):
    
    def __init__(self, s):
        self.s = s
        self.n = None
        self._complementer = None
        
    def __add__(self, other):
        if len(self)!=len(other):
            raise ValueError
        s = ''
        for i in range(len(self)):
            s += str((int(self.s[i])+int(other.s[i]))
        return OpString(s)
            
    def __sub__(self, other):
        if len(self)!=len(other):
            raise ValueError
        s = ''
        for i in range(len(self)):
            s += str((int(self.s[i])-int(other.s[i]))
        return OpString(s)
    
    def op_n(self):
        if self.n is None:
            n = 0
            for i, x in enumerate(reversed(self.s)):
                n += int(x)*3**i
            self.n = n
        return self.n
    
    def complement(self):
        if self._complementer is None:
            self._complementer = OpString("3"*len(self.s))
        return self._complementer - self
\end{python}

Finally, this \textit{OpString} class is integrated into the $\beta{(i)}$ parameter of our tuple-notation for context eigenstates, such that it becomes

\begin{equation}
\ket{\phi_{z\mbox{\tiny $(i)$},\alpha\mbox{\tiny $(i)$}},\mathcal{C}^{s\mbox{\tiny $(i)$}}_{\beta\mbox{\tiny $(i)$}}}\Rightarrow\bigg(\mathcal{C}^{(i)}_\text{ID}\,,\,\text{str}\Big(s{(i)}\Big)+\text{str}\Big(z{(i)}\Big)+\text{str}\Big(\text{binary}\big(\alpha{(i)}\big)\Big)\,,\,\text{OpString}\Big(\text{str}\big(\text{ternary}(\beta{(i)})\big)\Big)\bigg)\text{,}
\label{eq:state_tuple}
\end{equation}

\noindent where the first element is an integer, the second element is a regular string (of $0$s and $1$s), and the third element is an \textit{OpString} (string of $0$s, $1$s and $2$s with custom operations).

%% file: app-collective-parity-cost-function.tex
\section{Collective parity cost function}
\label{app:collective_parity_cost_function}

For any given event, each data bit $b_\gamma$, with $\gamma=0,1,\text{...},(3^n-1)/2-1$, is encoded as best as possible in an estimator $\mathcal{B}_\gamma$, obtained from the measurement outcomes of a pair of observables $\{\mathcal{O}_{2\gamma},\mathcal{O}_{2\gamma+1}\}$ over the state selection $\big\{\ket{\phi}\big\}$. This estimator is given by the collective parities of the pair of observables, which are added modulo-$2$ (XOR):

\begin{equation}
\mathcal{B}_{\gamma}={\rm{round}}\Big[\langle \mathcal{P}_{2 \gamma} \rangle\Big] \oplus {\rm{round}}\Big[\langle \mathcal{P}_{2 \gamma +1} \rangle\Big]\text{,}
\end{equation}

\noindent where

\begin{equation}
\langle \mathcal{P}_i \rangle=\text{bit}\bigg(\frac{1}{\left|\left\{\left|\phi\right>\right\}\right|}\sum_{j}\left<\phi_{j}\right|\mathcal{O}_{i}\left|\phi_{j}\right>\bigg)\text{.}
\end{equation}

\noindent Additionally, the round function and sum-modulo-$2$ are chosen to output $0.5$ if any input is $0.5$.

While the choice of encoding each bit in two observables is convenient to reduce the size of the selection, it increases the complexity of training the selection itself. Specifically, it is impossible to define a cost function for each individual context eigenstate without being aware of the rest of the selection, as the estimator $\mathcal{B}_\gamma$ depends on pairs of observables, whose eigenstates in the selection in general do not overlap. To solve this, we choose a desired collective parity $\mathfrak{b}_i$ for each observable $\mathcal{O}_i$ that is compatible with our data, that is, satisfies

\begin{equation}
b_\gamma=\mathfrak{b}_{2 \gamma}\oplus \mathfrak{b}_{2 \gamma+1}\text{.}
\label{eq:desired_parity_restriction}
\end{equation}

\noindent Since there are $3^n-1$ Pauli observables with a pair (one observable is left out), we have $i=0,1,\text{...},3^n-2$.

To choose appropriate values of $\mathfrak{b}_i$, we maximize their resemblance with existing eigenvalues in the contexts. For that end, let us index the contexts with Eq.~\eqref{eq:context_id} so as to identify them via 

\begin{equation}
\mathcal{C}_j \equiv \mathcal{C}^s_\beta \big| j=\beta+s \times 3^n\text{.}
\end{equation}

\noindent Then, let us define the Hamming distance between the desired parity bitstring $\mathfrak{b}$ and context eigenstate $\ket{\phi_{z,\alpha},\mathcal{C}_j}$ as

\begin{equation}
\mathcal{H}\Big(\mathfrak{b}\,,\,\ket{\phi_{z,\alpha},\mathcal{C}_j}\Big)=\sum_{i \big|\mathcal{O}_i \in \mathcal{C}_j}\mathfrak{b}_i \oplus \mathcal{P}\Big(\ket{\phi_{z,\alpha},\mathcal{C}_j},\mathcal{O}_i\Big)\text{,}
\end{equation}

\noindent such that only the relevant subset of the desired parity bitstring is considered. To obtain $\mathfrak{b}$, we choose values for it that satisfy Eq.~\eqref{eq:desired_parity_restriction} and minimize the cost function

\begin{equation}
\mathcal{L}_\mathfrak{b}=\sum_{j=1}^{2 \times 3^n}\text{Min}_{z,\alpha}\bigg[\mathcal{H}\Big(\mathfrak{b}\,,\,\ket{\phi_{z,\alpha},\mathcal{C}_j}\Big)\bigg]\text{,}
\end{equation}

\noindent where the minimal Hamming distance in each context is considered from among its eigenbasis. Computing this cost function requires an exponential number of steps in $n$, but can be efficiently done for $n=8$. Furthermore, the values of $z,\alpha$ which minimize the Hamming distance for each context are noted and stored in a separate variable called \textit{palette}, of length $2\times 3^n$, equal to the number of contexts.

In this manner, our state-selection training consists in defining which states of the palette to use in the selection. This is represented by an optimization over a bitstring of size $2\times 3^n$, where each bit indicates whether the corresponding palette state is in the selection or not. We train this bitstring to minimize the value of

\begin{equation}
\mathcal{L}_\phi=\sum_{i=0}^{3^n-2} {\rm{round}}\Big[\langle \mathcal{P}_{i} \rangle\Big] \oplus \mathfrak{b}_i\text{,}
\label{eq:selection_cost_function}
\end{equation}

\noindent by means of a genetic-algorithm, explained in the following section.

%% file: app-state-optimization.tex
\section{Optimization of State Selection}
\label{app:optimization}
We implemented an optimized a genetic algorithm (GA)  to carry out the optimization required to find the states that most closely match the selected representation of the input logical bit string.
The motivation behind the GA, which gives rise to its name, is in biology itself, or specifically, evolution by natural selection, which is based on three pillars, namely, (1) \textit{inheritance}: traits can be inherited in one generation from the previous one; (2) \textit{variation}: random variations occur in the genetic code when creating new individuals, and (3) \textit{competition}: individuals compete with each other and the one who is most fit is able to reproduce and pass on their traits to the next generation.
Note the importance of the genetic code here: it is the information that can be inherited, can vary, and determines the fitness of the individual.
This is precisely the reason why working with a GA is especially suitable in our case, since we can think of the individuals in the ``population'' as the bit strings, and by introducing variations and competition among the bit strings, we can select for the ones that are ``most fit.''
More precisely, we employ the cost function $\mathcal{L}_\phi$ in Eq.~\eqref{eq:selection_cost_function}, taking as input a bitstring $\boldsymbol{v}$ of the states of the palette which are present in the selection, and returning the corresponding value of $\mathcal{L}_\phi$, representing the distance between the encoded parities $\text{round}\big[\langle \mathcal{P}_i\rangle\big]$ and target parities $\mathfrak{b}_i$.
The fitness of bitstring $\boldsymbol{v}$ is determined by this function, and we refer to this bitstring $\boldsymbol{v}$ as our \textit{individual}.

The first step in the GA is to create the population of individuals.
This is done by picking bitstrings at random of the desired length ($2\times 3^n$).
Their fitnesses are subsequently evaluated and ranked. 
An iterative process then begins, whereby a new ``generation'' of individuals is created every time, based on the fitness of every bitstring.
Every new generation is created in two basic steps, which are repeated until we reach the total population number, which remains constant throughout the optimization.
These two steps are \textit{crossover} and \textit{mutation}:

\begin{enumerate}
\item In the crossover step two individuals $\boldsymbol{v}$ are selected at random from the current population, with probability of being picked proportional to their fitness $\mathcal{L}_\phi$. This weighting ensures that better individuals are more likely to pass to the next generation.
Once these two bit strings are selected, with probability $p_{\text{cross}},$ a crossover happens.
This crossover probability $p_{\text{cross}}$ is a meta-parameter of the algorithm, which we set to $0.95$.
The crossover is based precisely on the same process that occurs in chromosomes, whereby one segment of one chromosome is placed in the same location of the second chromosome, and vice versa.
Note that the point at which the crossover occurs, which determines the length of the exchanged segment, is selected at random.
This process is represented pictorially below.
\begin{center}
\includegraphics[scale=0.3]{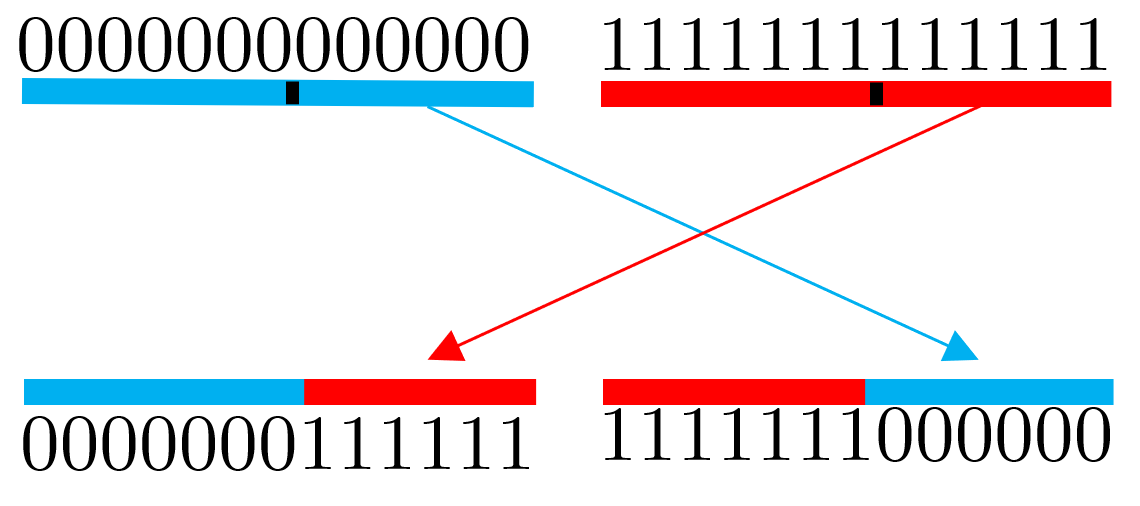}
\end{center}

\item In the mutation step, each bit of the individual is flipped with probability $p_{\text{mut}}$, where this value is another meta-parameter of the algorithm. Multiple bits may be flipped simultaneously, or a single one, or none.
\end{enumerate}

In this manner, via iterating these two steps, we obtain an individual $\boldsymbol{v}$ which represents a state selection with high resemblance to the desired collective parities $\mathfrak{b}$, which indirectly encode our original data $b$.